\documentclass[acmsmall]{acmart}

\AtBeginDocument{%
  }

\setcopyright{acmlicensed}
\copyrightyear{2018}
\acmYear{2018}
\acmDOI{XXXXXXX.XXXXXXX}

\usepackage{ulem}

\newcommand{\add}[1]{\textcolor{black}{#1}} 

\newcommand{\re}[1]{\textcolor{black}{#1}} 
\newcommand{\rev}[1]{\textcolor{black}{#1}} 

\acmConference[Conference acronym 'XX]{Make sure to enter the correct
  conference title from your rights confirmation email}{June 03--05,
  2018}{Woodstock, NY}
\acmISBN{978-1-4503-XXXX-X/18/06}

\usepackage{booktabs}
\usepackage{multirow}
\usepackage{graphicx}
\usepackage{xspace} 
\usepackage{natbib}
\usepackage{array}
\usepackage{tabularx}

\definecolor{cBLUE}{HTML}{3282B8}
\definecolor{cGREEN}{HTML}{60A561}
\definecolor{cORANGE}{HTML}{FA824C}
\definecolor{cYELLOW}{HTML}{f0C808}
\definecolor{cLightGrey}{HTML}{CECECE}
\definecolor{cRED}{HTML}{ED1B23}

\newcommand*{\etal}{\textit{et~al.}\xspace}

\begin{document}

\title{\add{Bridging Knowledge Gaps in Clinical AI: An Activity Theory Perspective on Interdisciplinary Data Work for Telehealth}}

\author{Bingsheng Yao}
\authornotemark[1]
\email{b.yao@northeastern.edu}
\affiliation{%
 \institution{Northeastern University}
 \city{Boston}
 \country{USA}
}

\author{Yao Du}
\authornote{Equal Contribution.}
\email{yaodu@usc.edu}
\affiliation{%
 \institution{University of Southern California}
 \city{Los Angeles}
 \country{USA}
}

\author{Yue Fu}
\email{chrisfu@uw.edu}
\affiliation{%
 \institution{University of Washington}
 \city{Seattle}
 \country{USA}
}

\author{Xuhai "Orson" Xu}
\email{xx2489@columbia.edu}
\affiliation{%
 \institution{Columbia University}
 \city{New York}
 \country{USA}
}

\author{Yanjun Gao}
\email{yanjun.gao@cuanschutz.edu}
\affiliation{%
 \institution{University of Colorado}
 \city{Aurora}
 \country{USA}
}

\author{Hong Yu}
\email{hong yu@uml.edu}
\affiliation{%
 \institution{University of Massachusetts Lowell}
 \city{Lowell}
 \country{USA}
}

\author{Dakuo Wang}
\authornote{Corresponding author.}
\email{d.wang@northeastern.edu}
\affiliation{%
 \institution{Northeastern University}
 \city{Boston}
 \country{USA}
}

\renewcommand{\shortauthors}{Yao et al.}

\begin{abstract}

Advanced AI technologies are increasingly integrated into clinical domains to advance patient care.
The design and development of clinical AI technologies necessitate seamless collaboration between clinical and technical experts. 
However, such interdisciplinary teams are often unsuccessful, with a lack of systematic analysis of collaboration barriers and coping strategies.
This work examines two clinical AI collaborations in the context of speech-language pathology via semi-structured interviews with six clinical and seven technical experts. 
Using Activity Theory (AT) as our analytical lens, \re{we examine persistent knowledge gaps and collaboration tensions across clinical and technical workflows, and show how clinical data can function as boundary objects while interdisciplinary collaborators may act as knowledge brokers to help address these challenges.}
\re{Our findings contribute to CSCW research on interdisciplinary teams' data work by showing how shared clinical data, boundary objects, and broker roles shape coordination in early-stage clinical AI collaboration, and by providing insights into best practices for future collaboration.}

\end{abstract}





\maketitle

\section{Introduction}

Recent advances in artificial intelligence (AI) open up a broad avenue of interdisciplinary team collaboration opportunities, especially in the clinical and health domain~\cite{yamazaki2023exploration, janowski2023natural, li2024academic, yamazaki2023exploration, chen2020trends, jiang2023health}.
Early clinical AI collaborations predominantly focus on the analysis of clinical information from electronic medical records (EMR) systems to support both administrative (e.g., documenting patient admission and organizational data)~\cite{park2012adaptation, gui2020physician, pine2022investigating, pine2023innovations} and clinical inquiries (e.g., disease progression modeling, treatment effectiveness analytics) in hospital settings~\cite{chen2020trends, helou2019understanding, linhares2022clinicalpath, cui2022automed}.
More recent work extends AI-driven capabilities into ubiquitous computing contexts, where patient-provider communication~\cite{yang2024talk2care, wu2024clinical}, clinical decision-making~\cite{jiang2023health, xu2024mental, osong2025development}, remote monitoring~\cite {shaik2023remote, li2025nerve}, as well as assessment~\cite{yang2025recover, yao2025more, wu2024cardioai}, are increasingly supported by advanced AI technologies.
\add{The successful design and development of these applications depends on the close collaboration between clinical and technical experts to produce outcomes that align with real-world healthcare practices.}

\add{Interdisciplinary work in healthcare has been extensively researched by HCI/CSCW researchers.
Foundational studies have examined the complexities of coordinating clinical tasks and the significant effort involved in "articulation work," which includes the invisible labor required for managing dependencies and aligning disparate activities in complex scenarios~\cite{schmidt1992taking, schmidt1994cooperative}.
While clinicians' workflow strictly follows established institutional guidelines and prioritizes patient outcomes, technical experts are primarily data-driven, focusing on well-defined computational tasks and optimizing for model performance metrics~\cite{hernandez2020minimar, nagendran2020artificial}.
Such misalignment in domain-specific goals and practices can lead to fundamental disagreements about collaboration goals, distribution of labor, and even the definition of a successful outcome~\cite{singh2017hci, bossen2019data, pine2022investigating}. 
Although these obstacles are noted in a scatter of individual case studies and surveys \cite{guhan2025developer, shamszare2023clinicians}, current research lacks a deeper analysis regarding the underlying structural conflicts and tensions that commonly emerge between clinical and technical teams, and how interdisciplinary teams cooperatively mitigate these challenges when integrating AI into the complex clinical contexts. Understanding the process of work across disciplinary boundaries can help CSCW researchers, as well as interaction designers and developers for health information technology, to better communicate and collaborate when creating AI technologies for health.}

\add{To bridge the gap, the present study leverages \textbf{Activity Theory (AT)} as a guiding analytical lens to identify challenges and coping strategies between clinical and technical experts in Clinical AI collaborations~\cite{engestrom2000activity, nardi1996context}.}
AT is a conceptual framework for understanding human activities and their social, cultural, and historical contexts through a unified lens which considers human activity as a \add{system of interacting elements: the subject (the individual or group), the object, tools, rules, community, and the division of labor.}
In the context of healthcare, AT has been productively adopted to investigate and examine how clinical tools and procedures shape collaborative tasks~\cite{durst2017guideline, clemmensen2016making, engestrom2018expertise, fossouo2023linking, grundgeiger2024motives}.
\add{AT is particularly suitable for our investigation of the interdisciplinary Clinical AI collaboration due to several key reasons. 
First, since AT can describe hierarchical structures and decompose actions and operations of activities \cite{hashim2007activity}, using AT as an analytical framework provides a systematic view to model the distinct but interconnected activities (e.g., data analysis/interpretation, documentation, and organization) between clinical and technical experts. 
Second, leveraging AT as an analytical framework enables researchers to dive deep into the fundamental misalignment or conflicts in objects, tools, or rules underneath the surface-level communication and coordination difficulties.
Third, since AT recognizes that activities are mediated by tools, such an epistemological stance helps researchers to explain relationships between the user and the tool in interdisciplinary collaboration contexts. 
Collectively, AT becomes a promising framework to help identify best practices for interdisciplinary research and collaboration in Clinical AI.}

Building on these insights, we applied AT to analyze two cases of interdisciplinary collaboration between speech-language pathologists (SLPs) and technical experts who are applying AI approaches such as natural language processing (NLP) to optimize clinical workflows in the telehealth context via remote research collaborations.
Through semi-structured interviews with a total of 13 clinicians (N=6) and technical experts (N=7) from these collaborations, we investigated how both groups \add{navigate the collection, coding, and transformation of clinical data}
for downstream AI pipelines. 
Our analysis was guided by the following research questions:
\begin{itemize}
    \item RQ1: How do clinicians and technical professionals perform their workflows independently and collaboratively on clinical AI research?
    \item \add{RQ2: What \re{communication tensions} and data-specific challenges arise and how do interdisciplinary teams develop coping strategies to navigate and resolve these challenges?}
    
\end{itemize}

\add{Our analysis highlights how AT can be utilized as an analytical framework to systematically 1) unpack how interdisciplinary knowledge workers utilize different data tools to perform domain-specific workflows (e.g., clinical and technical), 2) reveal the tension points in distinct yet intertwined workflows between different groups of experts, and 3) identify communication barriers arise and corresponding coping strategies to resolve these challenges. Specifically, our findings show that concrete clinical data often act as "boundary objects" \cite{star1989institutional} that anchor the discussion, while human knowledge brokers with cross-disciplinary expertise guide critical activities of ``bi-lingual translations'' of specialized languages in Clinical AI collaboration. \re{It further identifies core conflicts (e.g., knowledge gaps in taxonomy) grounded in the divergent professional practices (e.g., different handling of data) and goals (e.g., clinical documentation vs. model deployment), which become most visible in the misaligned work around clinical data.} By systematically answering research questions through the lens of AT, our contributions broaden current CSCW and HCI knowledge of how complex clinical and technical workflows integrate into the broader ecosystem of AI development for real-world, domain-specific scenarios. Through the lens of AT, we demonstrate how focusing on shared objectives, mediating artifacts, and proper labor division following disciplinary rules can illuminate complex collaboration patterns. This perspective offers practical guidance for future HCI/CSCW research and system design aimed at empowering AI solutions to support interdisciplinary team collaborations.}
The findings deepen our understanding of how distinct bodies of expertise can jointly produce AI-enabled solutions while managing competing terminologies, professional identities, and research demands.
In light of rapid technological advances, recognizing and addressing these nuanced interdisciplinary dynamics is the key to ensuring robust, trustworthy, and effective human-centered AI in team collaboration.

\section{Related Work}

\subsection{Activity Theory as Analytical Frameworks in Exploring Interdisciplinary Team Collaboration}

\begin{figure}[!tp]
    \centering
    \includegraphics[width=.9\linewidth]{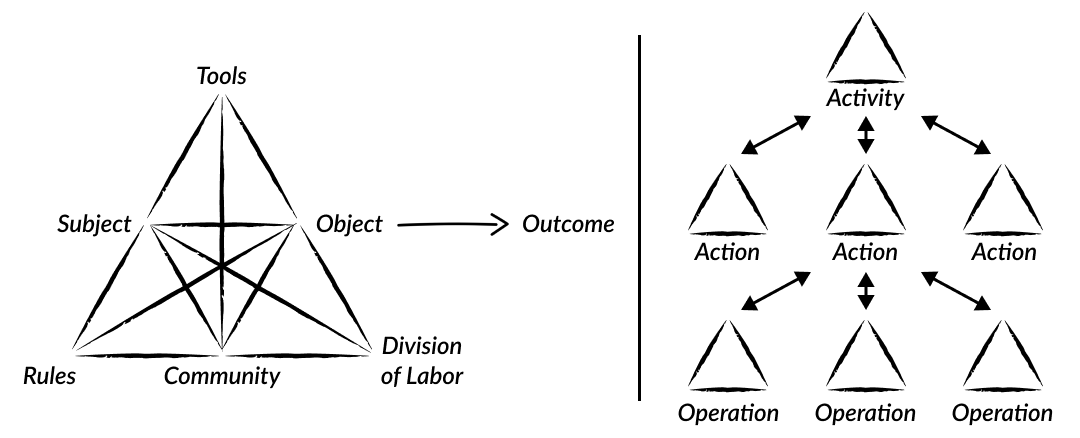}
    \caption{Left: Engeström's triangle of Activity Theory~\cite{engestrom2015learning}, also known as ``Activity Triangle''. Right: the hierarchical structure of activities~\cite{kaptelinin2009acting}.
    }
    \Description{Engeström's Triangle of Activity Theory, also known as ``Activity Triangle''. In this work, we primarily focus on the top triangle in the AT system, which constitutes the subject, object, instrument, and the relations between. }
    \label{fig:activity-triangle}
\end{figure}

AT offers a comprehensive framework for understanding the complex interplay of human activities, tools, and social contexts in collaborative settings.
Originally developed from the work of Vygotsky and later expanded by Engestrom, AT provides a lens through which interdisciplinary team collaborations can be analyzed by focusing on how people, tools, and broader organizational structures interact by using an activity triangle~\cite{engestrom2000activity, durst2017guideline, clemmensen2016making}.
As shown in Figure~\ref{fig:activity-triangle} (left), a typical activity triangle proposed by \citet{engestrom2000activity} includes seven key elements that constitute a comprehensive view of an individual activity, including subject, object, instrument, rules, community, division of labor, and outcome.
In particular, the most critical elements in an activity triangle are the top triangle that focuses on the relationships and interactions between subject, object, and instrument, where \citet{engestrom2000activity} complements the diagram with the addition of the bottom three elements.
In this study, we primarily utilize the top triangle to understand the activity systems of both clinical and technical experts' workflows, while also leveraging the complete AT triangle to conduct a holistic analysis of misalignment and conflicts in Clinical AI collaboration.
In addition, \citet{kaptelinin2009acting} presents a hierarchical structure of activity by decomposing an activity into higher granularities and aligning with different levels of psychological notions as the target outcome. 
As shown in Figure~\ref{fig:activity-triangle} (right), \add{the hierarchical AT structure distinguishes between the motive-driven activity, the goal-driven actions that constitute it, and the condition-driven operations, which allows researchers to deconstruct complex workflows and identify systemic contradictions that may not be immediately apparent.}


\add{Reviews in HCI explain how AT functions as an analytic and design framework and clarify common ways it is used in empirical studies \cite{clemmensen2016making}.
Particularly, AT has been adopted in several distinct ways to study complex healthcare contexts.
In CSCW, studies used AT to analyze hospital work and derive design implications, including activity analysis of ward and operating room workflows and the activity-based computing to guide clinical infrastructures that support multitasking, mobility, and coordination in clinical settings~\cite{bardram2009activity, bardram2011activity}.
In addition, AT has been used to define the object of work and identify challenges that inform the design and evaluation of specific health technologies, such as understanding patients' activities in self-quantification~\cite{almalki2016activity}, managing risk around patient health information~\cite{valecha2021activity}, and designing adverse drug reaction reporting tools~\cite{fossouo2023linking}.
AT also served as a powerful lens for understanding and facilitating learning and organizational change, such as examining how medical teams develop expertise and innovate their practices~\cite{engestrom2018expertise, kerosuo2010promoting}.
Despite this extensive body of work, most studies focus on situated clinical work and documentation, but few examine end-to-end Clinical AI development as a joint activity across data work, modeling, evaluation, and integration. 
Our work addresses this gap by using AT to examine where objects, tools, rules, and division of labor misalign or conflict, and what mediating strategies are adopted and used by interdisciplinary clinical vs. AI teams.}
Understanding these dynamics can inform the development of tools and strategies to better support interdisciplinary teams in effectively collaborating and achieving shared objectives.

\subsection{Knowledge Brokers and Boundary Objects in Interdisciplinary Team Collaboration}

A longstanding challenge in interdisciplinary team collaboration is managing \textbf{knowledge boundaries} that separate expert fields, such as clinical and technical domains.
Knowledge boundaries arise due to differences in education, professional training, terminology, and methodological approaches, which can lead to communication difficulties and misunderstandings. 
Boundary objects~\cite{star1989institutional} and knowledge brokers~\cite{kimble2010innovation} play a critical role in facilitating communication and collaboration between knowledge boundaries. 
Specifically, \textbf{boundary objects} are defined as ``plastic and robust'' entities that can be interpreted differently by each group while maintaining a common structure sufficient to be recognized and used by all parties involved~\cite{bowker1999sorting}. 
Boundary objects are critically valuable in interdisciplinary research because they can help bridge the cognitive, methodological, and epistemological gaps between disciplines~\cite{wilson2007boundary, akkerman2011boundary, leigh2010not}. 
For example, in large-scale collaborations, boundary objects like data models, frameworks, and tools enable stakeholders to work together effectively despite having distinct perspectives, knowledge, and goals \cite{akkerman2011boundary}.

In clinical AI, the distinct educational and training background between clinical and AI experts results in communication and collaboration difficulties because of the different knowledge, language, and skills, as well as different priorities of documentation and use of clinical data~\cite{janowski2023natural, singh2017hci, bossen2019data}. 
Clinicians typically use Electronic Medical Records (EMRs) and other non-EMR tools, such as spreadsheets, to support convenient access, retrieval, and analysis of structured patient data for better clinical decision-making and patient outcome~\cite{juluru2015use, taylor2020research, thorne2017role, iyengar2020big, van2017boundary}. 
Carlile \etal~\cite{carlile2002pragmatic, carlile2004transferring} emphasizes the role of boundary objects in facilitating knowledge sharing across different domains, highlighting the processes of transferring, translating and transforming knowledge to overcome communication barriers. 
\add{Zhou \etal~\cite{zhou2011cpoe} explore how Computerized Physician Order Entry (CPOE) systems act as boundary objects and how clinicians developed workarounds to bridge the gaps between different work practices and technologies in healthcare settings.}
Despite their prevalence, the role of non-EMR tools as boundary objects in clinical AI research remains underexplored. 
Spreadsheets, for instance, are flexible and easily customizable, making them valuable tools for data entry, analysis, and sharing in various clinical and research settings \cite{ponedal2002understanding, thorne2017role, iyengar2020big, janowski2023natural, demir2012decision}.
Clinicians can use such tools to encode patient data and document their insights, where documented data can function as mediators in knowledge exchange between clinicians and AI experts.

\textbf{Knowledge brokers} are closely related to boundary objects in terms of facilitating the exchange, translation, and application of knowledge between different professional communities \cite{lomas2007between, kimble2010innovation}. 
For instance, IT professionals often serve as knowledge brokers to facilitate knowledge sharing and integration across organizational boundaries through their technical expertise in organizational contexts \cite{pawlowski2004bridging}. 
In the context of science communication, knowledge brokers help ensure that scientific research effectively informs practice and decision making~\cite{meyer2010rise}.
By enabling better communication, collaboration, and innovation, both boundary objects and knowledge brokers are essential for successful cross-boundary collaboration and innovation \cite{caccamo2023boundary}, \add{and are also evident in clinical experts who worked as knowledge brokers across different disciplines of care \cite{cross2023roles, douglas2022knowledge, mickan2022using, gaid2023barriers, yamanie2023impact}.}

In clinical AI, clinical and technical experts may use different terminologies to describe concepts such as "model," which will lead to different ways of collection, management, organization, and analysis of clinical data when attempting to apply AI approaches in a collaborative and interdisciplinary research context. Despite the fruitful study results reporting challenges in interdisciplinary team collaboration in medical and healthcare scenarios~\cite{cross2023roles, douglas2022knowledge, mickan2022using, gaid2023barriers, yamanie2023impact}, there is a lack of work in the Clinical AI context, which could be attributed to the relatively new and rapid development of both expert fields in recent years.
Thus, this work fills the gap by understanding how clinicians and AI experts perform and collaborate, as well as analyzing the needs and challenges of such interdisciplinary team collaborations.



\subsection{Interdisciplinary Team Collaboration in Clinical AI Research}


By integrating knowledge, expertise, and perspectives across disciplines, interdisciplinary teams can support innovation in complex problem spaces~\cite{schmidt2008towards, barry2008logics, blackwell2009radical, wu2019large}. \re{In healthcare, interdisciplinary collaboration has long been central to CSCW research. }
Prior work has shown the importance of understanding the dynamic and diverse needs of clinical professionals and translating those needs into functional requirements for information systems, rather than imposing one-size-fits-all solutions~\cite{reddy2001coordinating}. 
Successful clinical technologies, therefore, require careful attention to human factors and close collaboration among clinicians, engineers, HCI specialists, and other stakeholders to improve both technology design and patient outcomes~\cite{bednarik2022integration}.

The recent expansion of AI in healthcare has introduced new demands for interdisciplinary collaboration, often requiring teams from different fields to work together on complex clinical problems~\cite{agapie2024conducting}. 
For instance, Lyon et al.~\cite{lyon2023bridging} discuss the importance of integrating HCI with implementation science to improve the adoption, implementation, and long-term use of innovations in public health. 
However, such collaborations also introduce distinct challenges. 
Experts from different backgrounds bring different perspectives, priorities, and vocabularies, which can create knowledge gaps and communication barriers~\cite{mozgai2024accelerating}. 
To address these difficulties, Nagendran et al.~\cite{mozgai2024accelerating} propose user-centered design workshops as one mediating strategy. 
In another interview study with 20 academic and industry AI developers, Guhan et al.~\cite{guhan2025developer} examined how AI-based computer perception technologies were developed for clinical decision-making. 
\re{Their participants reported that this work not only supported existing clinical workflows but also prompted clinicians to consider alternative perspectives on illness and care.}

\re{While prior studies have shown that interdisciplinary clinical AI collaboration can generate innovation across clinical and technical domains, we still know less about how clinicians and technical experts coordinate around shared data in day-to-day collaborative work, and where tensions emerge in that process. 
Our work addresses this gap through two case studies of clinical AI collaboration, using Activity Theory to analyze how data work, communication barriers, and mediating practices shape interdisciplinary coordination.}

\section{Method}

This study employed a case-based, semi-structured interview methodology to examine interdisciplinary collaborations between clinical experts and technical experts, particularly in the context of remote research collaborations on projects in telehealth for both children and adults.
We recruit both groups of experts from two team collaborations between speech-language pathologists (SLPs) and technical experts with AI expertise, where both groups of experts are from academic research backgrounds.
Although each collaboration addressed distinct clinical challenges, both followed a similar model of interdisciplinary team collaboration: Clinical experts collaborate with technical experts to empower advanced AI models that could be beneficial to the clinical practices of SLPs.

\subsection{Study Context and Case Selection}

We selected these two collaborations as case studies due to the following shared commonalities. 
Both cases are related to the field of speech-language pathology settings and focused on telehealth scenarios. 
\re{Both projects shared two primary objectives for the interdisciplinary teams: improving patient care through AI-assisted telehealth assessment or intervention, and exploring how clinicians could use AI-driven insights to support clinical decision-making and reduce clinical workload.}
The teams work remotely and rely on Zoom meetings throughout the collaboration period, which provides an excellent context for us to study cross-disciplinary collaboration. 
These shared characteristics allowed us to investigate common interdisciplinary workflow patterns without becoming overly confined to a single project’s particularities. Still, each project has unique patient care needs and individual differences in clinical tasks that lead to different types of communication and collaborations with AI experts. Case 1 involves bilingual clinicians who work with bilingual child language assessment data, whereas Case 2 involves monolingual clinicians who work with adults with neurogenic disorders. Also, both clinical and technical research teams are affiliated with different academic institutions in the United States. 
\re{For our study,} we centered our inquiry on how participants conceptualized, documented, and processed patient data to develop or refine AI tools for clinical use and how they communicated across disciplinary boundaries.


\subsubsection{\add{Case 1: Tele-Assessment for Bilingual Children}}

\add{Conventional language assessment for bilingual children is a clinical challenge due to a shortage of bilingual resources and professionals, often leading to inaccurate diagnoses~\cite{du2020try}.
This Clinical AI project aims to develop a web-based telehealth tool that allows parents to supervise automated language assessments for their children at home. 
While the tool facilitates data collection, clinicians must still perform intensive manual labor to score, analyze, and interpret video data for diagnostic decision making. The primary goal of this project is to \textbf{use AI to automate clinical data analysis}, with the aim of reducing clinicians' workflow burden and efficiently identifying parent-child dyadic behaviors and communication. The collaboration focuses on developing multimodal AI pipelines to support clinicians, bilingual children, and their bilingual parents.}

\subsubsection{\add{Case 2: Telehealth Training for Adults with Brain Injury}}

\add{Individuals with cognitive-communication disorders (e.g., traumatic brain injury, aphasia, or dementia) often use voice assistive technologies (e.g., Amazon Alexa) to support their performance of daily activities, such as medication management~\cite{du2024voice}. However, due to their cognitive-communication difficulties, these patients often need additional training and exercises from clinicians to learn how to use these assistive technology tools effectively. The primary goal of this project is to \textbf{explore how AI can enhance the efficiency and quality of this telehealth training}, for instance, by streamlining the customization of voice commands for individual patients. The collaboration focuses on developing AI-driven methods to support clinicians, adult patients, and their caregivers.}

\subsection{Participants and Recruitment}

To gain comprehensive insights into interdisciplinary team collaboration between clinical and technical experts across the two projects, we employed a convenience sampling strategy for participant recruitment.
We recruited a total of 13 individuals \add{(Males: N=3, Females: N=10)} involved in \add{two case study collaborations (Case 1: N=7, Case 2: N=6)} to participate in semi-structured interviews, where the demographics of participants are reported in Table~\ref{tab:participants}. 
\add{To ensure balanced role-case distribution and representative findings for both case studies, participants include six SLPs (Case 1 N=3, Case 2 N=3) and seven technical experts (Case 1: N=4, Case 2: N=3). All clinicians hold at least a master’s degree in speech-language pathology and actively interact with patients to collect and analyze data for research; all technical experts have at least a bachelor's degree and play active roles in designing or implementing AI approaches using data obtained from clinical experts.} All participants were actively contributing to one of the two collaborations, ensuring they had recent, first-hand insights into clinical–technical interactions, data-sharing practices, and communication challenges.

\begin{table}[t]
\footnotesize
\caption{Demographic information of participants \add{(Case 1 N=7, Case 2 N=6; Clinicians N=6, AI Experts N=7)}.  }
\resizebox{.95\textwidth}{!}{%

\begin{tabular}{cccccc}
\toprule
Participant & Gender & Education & Expertise & \begin{tabular}[c]{@{}c@{}}Year of Experience \\ (AI/Clinical) \end{tabular} & \add{Case Study} \\ 
\midrule 

P1  & F & Master   & Clinician       & <1 year  & \add{Case 2} \\
P2  & F & Master   & Clinician       & <1 year  & \add{Case 2} \\
P3  & F & Master   & Clinician       & <1 year  & \add{Case 2}\\
P4  & F & Master   & Clinician       & 1-5 years & \add{Case 1}\\
P5  & F & Master  & Clinician       & 1-5 years & \add{Case 1}\\
P6  & M & Doctorate & AI Expert & 1-5 years & \add{Case 2}\\ 
P7 & F & Doctorate & AI Expert & <1 year   & \add{Case 1}\\
P8 & F & Doctorate & AI Expert & 1-5 years & \add{Case 1}\\
P9 & M & Bachelor & AI Expert & <1 year   & \add{Case 1}\\
P10 & F & Bachelor & AI Expert & 1-5 years  & \add{Case 2}\\
P11 & M & Master & AI Expert & 1-5 years   & \add{Case 1}\\
P12 & F & Doctorate & Clinician & 1-5 years  & \add{Case 1} \\
P13 & F & Doctorate & AI Expert & 5+ years  & \add{Case 2}\\ 

\bottomrule
\end{tabular}%
}

\label{tab:participants}
\end{table}

\subsection{Interview Development \& Data Collection}

Given the complicated nature of interdisciplinary research that involves both independent and collaborative work across diverse fields, we employed the AT framework to structure the development of interview questions. 
In this study, the AT framework helps deconstruct the complexities inherent in interdisciplinary collaborations between clinical and technical teams, highlighting contradictions and tensions within project workflows. The interview protocol \add{(Appendix~\ref{app:protocol})} was crafted to capture the dynamics of workflows among clinical and technical experts, their experiences with clinical and technical-related tools, and the challenges they face in collaboration and communication during interdisciplinary projects. To ensure relevance and specificity, the interview questions were tailored to the unique roles and experiences of each professional group, with respect to the AT framework. 
\add{For instance, we first asked clinicians about their objectives for patient care and technical experts about their goals for AI model performance to understand the \textbf{Object (goal)} of the activity.
We then asked all participants to describe the software, spreadsheets, and protocols they used and to share their screens to show examples to investigate the \textbf{Tools} used in various clinical activities.
We gathered information about formal guidelines (e.g., clinical standards) and informal "unwritten rules" that shaped clinicians' work to probe the \textbf{Rules} and norms (e.g., best practices within their disciplines). We also asked participants to describe who was responsible for specific tasks (e.g., data collection, cleaning, labeling, and modeling) and where handoffs occurred to map the \textbf{Division of Labor}.
Lastly, we specifically asked about challenges, frustrations, or breakdowns that occurred during data interpretation or handoffs to identify misalignment and conflicts across disciplines. }

All 13 clinical and technical experts from the two collaborations completed an electronic consent form for demographic information through a Qualtrics survey. 
The interviews lasted between 35 and 60 minutes and were video-recorded via Zoom with participant consent. In addition to answering interview questions, participants were encouraged to refer to their data by providing descriptive examples of their clinical or technical work via Zoom's screen-sharing feature. 
This approach facilitated better recall and explanation of specific clinical or technical activities accomplished via tools such as spreadsheets. 
This study was approved by the Institutional Review Board of the affiliated institutions of the researchers involved in this human subject research.

\subsection{Data Analysis}

\add{All interviews were transcribed verbatim and analyzed by two researchers (first and second author) with corresponding AI and clinical backgrounds to ensure unbiased interpretation across two disciplines. Two phases of analysis were utilized in interpreting the interview transcripts: qualitative analysis via inductive coding and thematic analysis \cite{saldana2021coding, braun2006using, braun2019reflecting}, followed by research diagramming using AT as an analytical framework \cite{chigbu2019visually, crilly2006using}. First, using inductive coding \cite{saldana2021coding, creswell2017designing}, the first author (an AI expert) coded a subset of 8 transcripts (P1-P8, Case 1 N=4, Case 2 N=4) using open coding, then grouped codes to identify key themes while iteratively refined the coding with a focus on various clinical and technical activities. Next, based on prior AT research in health \cite{engestrom2018expertise, bharosa2012activity, valecha2021activity, fossouo2023linking, grundgeiger2024motives}, the first author worked with the second author (a clinical expert who coded the same set of transcript) and developed a qualitative codebook using existing constructs in the AT framework as individual units of analysis (Appendix~\ref{app:protocol}).}

\add{\re{Next, using thematic analysis} \cite{braun2006using, braun2019reflecting, fereday2006demonstrating}, key themes (especially related to clinical and technical activities) that emerged from the transcripts were reviewed and refined iteratively to ensure that they accurately represented the data. To ensure consistency and quality in coding, both coders discussed discrepancies during coding, and resolved disagreement through consensus in the qualitative codebook. Both researchers reviewed final themes in the context of the study's research questions, with particular attention paid to (1) the relationship between participants and their diverse approaches to patient data during interdisciplinary team collaboration for both projects, and (2) the different tools used as "boundary objects" and their potential to facilitate or hinder interdisciplinary team collaboration.}

\add{Finally, to integrate AT as an analytical framework, using research diagrams \cite{crilly2006using, chigbu2019visually}, both authors created detailed illustrations to visualize elements of the activity triangle (e.g., subjects, objects, and tools) with contextualized representations (clinical vs. technical) based on two case studies. Every research diagram is annotated with critical relations between different AT entities, offering implications to reveal underlying tensions or challenges reported during the interviews. To ensure consistency and quality in research diagramming, both coders discussed discrepancies in diagramming and resolved disagreements through consensus in the research diagrams. After this, the remaining five transcripts were transcribed and analyzed. It is important to note that using deductive and inductive approaches to analyze transcripts enables authors to provide both bottom-up and top-down interpretations for the interview data. Using this approach ensures that findings can converge on key constructs from the AT framework (e.g., use of data-related "Tools") across two disciplines, while enabling divergent interpretation on other themes (e.g., clinical vs. technical workflow, communication conflicts) during interdisciplinary team collaborations. This approach also enabled researchers to analyze the mediating strategies and solutions that participants employed to communicate and collaborate with each other, providing insight into the complex dynamics of interdisciplinary collaborations within and between the clinical and NLP teams.}


\re{All interviews were transcribed verbatim. Two researchers, the first and second authors, analyzed the transcripts. They hold backgrounds in artificial intelligence and clinical practice, respectively, which supported a cross-disciplinary interpretation of the data. Our analysis proceeded in two stages. We began with inductive qualitative coding and thematic analysis, followed by research diagramming informed by Activity Theory~\cite{chigbu2019visually, crilly2006using}.}

First, the first author used open coding on a subset of transcripts~\cite{saldana2021coding, creswell2017designing}. This subset included four participants from Case 1 and four from Case 2. The author grouped these initial codes into preliminary themes related to clinical and technical activities. The second author coded the same subset of transcripts. The two authors then compared their interpretations and refined the developing codebook through discussion.

We subsequently reviewed and refined the emergent themes using thematic analysis~\cite{braun2006using, braun2019reflecting, fereday2006demonstrating}. We paid particular attention to how participants described patient data, cross-role collaboration, and the artifacts used to support coordination. To maintain consistency, both coders discussed discrepancies and resolved disagreements through consensus. After stabilizing the codebook and preliminary themes, we coded the remaining five transcripts using the same process.

Activity Theory informed a later stage of the analysis. Both authors created research diagrams to visualize elements of the activity triangle~\cite{crilly2006using, chigbu2019visually}. We used these diagrams to compare the two cases and to interpret recurring tensions in how clinical and technical work was coordinated around shared data artifacts. This combination of inductive coding and AT-informed interpretation allowed us to identify themes grounded in participants' accounts while also examining broader patterns of interdisciplinary coordination.






\section{Findings}
\label{sec:findings}

\re{In this section, we present findings from our interviews with clinical and technical experts collaborating on two clinical artificial intelligence projects. We first describe how clinicians and technical experts engaged in distinct yet interdependent forms of data work. We then examine the communication challenges that arose as these groups coordinated across different terminologies, taxonomies, and workflow expectations, as well as the artifacts and broker roles that helped them manage these tensions.}

\add{\subsection{Using AT Analysis to Understand Clinical and Technical Workflow}}


\add{AT offers a structured and systematic lens for mapping the interactions, objects, and motives that shape these collaborations beyond identifying surface-level difficulties. Each AT triangle uniformly forms the ``core'' relationship between the subject, object, and tool. AT can also reflect how clinical and technical actions intersect, what contradictions or tensions arise, and how teams navigate them in the broader sociocultural context with respect to community, rules, and division of labor. Our qualitative analysis \re{shows} that while data anchors clinical and computational activities, the collaboration challenges between clinical experts and technical experts often extend beyond surface-level differences. We explore the ways clinical and technical experts work with patient clinical data in distinct yet overlapping processes throughout their workflows. Clinical specifications vary significantly across different specializations (educational pediatrics assessment vs. adult medical rehabilitation \textit{communities}), such as the clinical experts' training experiences, their daily workflows, and interactions with different patient populations. On the other hand, technical experts in team collaboration receive clinical data from their clinical partners and then create or adapt a technical taxonomy for computational purposes. }


\subsubsection{Clinical Workflow: Capturing the Nuances of Patient Care via Clinical Data}

Speech-language pathologists (SLPs) in our study typically begin by collecting patient data during face-to-face or remote telehealth sessions. One participant noted that remote sessions became increasingly common during and after the COVID-19 pandemic. \add{P2 (Case 2) explained the difficulty in understanding the clinical context when they are not able to physically present at the clinical sessions: \textit{“The subjective part is hard to understand when you are not present in during the session.”} These clinician reports showed that nuanced clinical knowledge, which guides how raw patient data are labeled, may not be obvious to collaborators on the technical side.} In these cases, patients interact with telehealth platforms, and various patient data are recorded during the interaction, including audio and video recordings, transcripts of patient verbal communications, and interaction history with telehealth platforms. A critical step in clinicians' workflow is to develop or adopt a clinical data taxonomy that represents each patient’s symptoms, behaviors, or responses into meaningful and useful categories.
Such taxonomy must follow domain-specific \textit{rules}, such as specifications and guidelines. \add{One clinician (P12, Case 1) explained the clinical workflow and documentation for bilingual child language assessment as the following:}

\begin{quote}
\textit{\add{“While parents/children talking with each other, this kind of interaction is hardest part (before or after the session) how long before the session/after the session you should start/stop documenting in the spreadsheet. Sometime it is important, but some of those are irrelevant, the timestamp is very important to mark.”}}
\end{quote}

The \textit{motive} to collect and encode patient data with a well-defined taxonomy centered on ensuring future clinical decisions are aligned with professional standards and medical guidelines, as well as maintaining consistency in patient encounters. 
For instance, SLPs frequently code children's behaviors and speech patterns during remote sessions with specialized taxonomies that reflect professional standards for communication disorders. 
These taxonomies guide how clinicians interpret and document different types of speech errors and relevant contextual factors. \add{Clinician P1 (Case 2) described the goal (motive) for teaching traumatic brain injury (TBI) patients smart home technology (e.g., Alexa) use at home by explaining the role and the clinical activities of an SLP:}

\begin{quote}
\textit{\add{“So when we're trying to help facilitate the communication and the clear speech, when trying to get them to execute the command, like, we'll use certain cues, such as like verbal prompting, visual prompting, or we may try to use some behavioral methods, like getting them to slow their speech down, or getting them to what's it called, use certain articulatory movements to get them to say the command more clear.”}} \end{quote}

\add{ P2 (Case 2) also described her clinical process of using telehealth when working with her patients based on a clinical protocol and documenting on the spreadsheet:}

\begin{figure}[!tp]
    \centering
    \includegraphics[width=0.95\linewidth]{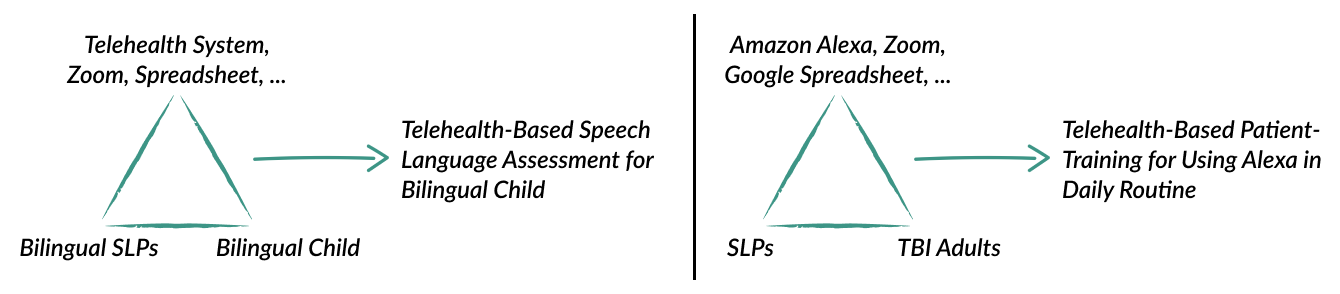}
    \caption{\add{Activity triangle for clinical workflow for the two case studies (Left: Case 1, Right: Case 2)}}
    \label{fig:activity-system}
\end{figure}

\begin{quote}
\textit{\add{“For the user needs assessment, when we start we have a template of questions that we usually go \re{by}. We have been going through each week based on that particular topic which \re{provides} enough guidance \re{for} that kind of template. And then the the training part, we have the spreadsheet with the columns for the accuracy of responses and the type of command, the prompts needed, etc. So we have we would record all of that on there.”} } \end{quote}

Once data have been collected, clinicians transform the raw materials into usable clinical documents or records. They code each observation in spreadsheets or specialized software such as electronic health record (EHR) systems, verify specific symptoms in annotated video clips, and summarize key findings in patient reports. These data transformations are motivated to help clinicians plan interventions or diagnose potential disorders and must comply with institutional guidelines and medical regulations. However, the complexity and domain specificity of these annotations can pose challenges to interpretation by collaborators within and outside the clinical domain. \add{Clinician P5 (Case 1) described collaborative processes working on bilingual data with another clinician:}

\begin{quote}
\textit{\add{“I work with another fellow person and we talked about each definition together, whether this be in Mandarin or English. Did it happen in Mandarin? Did it happen in English? Just collaborating with this person \re{who} was also another speech (therapy) person and making sure that we were in cohesive definition of from the codebook, codebook's very important.”}} \end{quote}

In summary, using AT analysis for visual illustration, SLPs are the \textit{subject} by \add{leveraging professional practice protocols, guidelines, and telehealth systems (the \textit{tools}) to transform raw observations with patients (\textit{object}) into meaningful clinical data documentations, as described in Figure \ref{fig:activity-system}.}


\subsubsection{Technical Workflow: Preparing Data for Model Pipelines}

\begin{figure}
    \centering
    \includegraphics[width=0.6\linewidth]{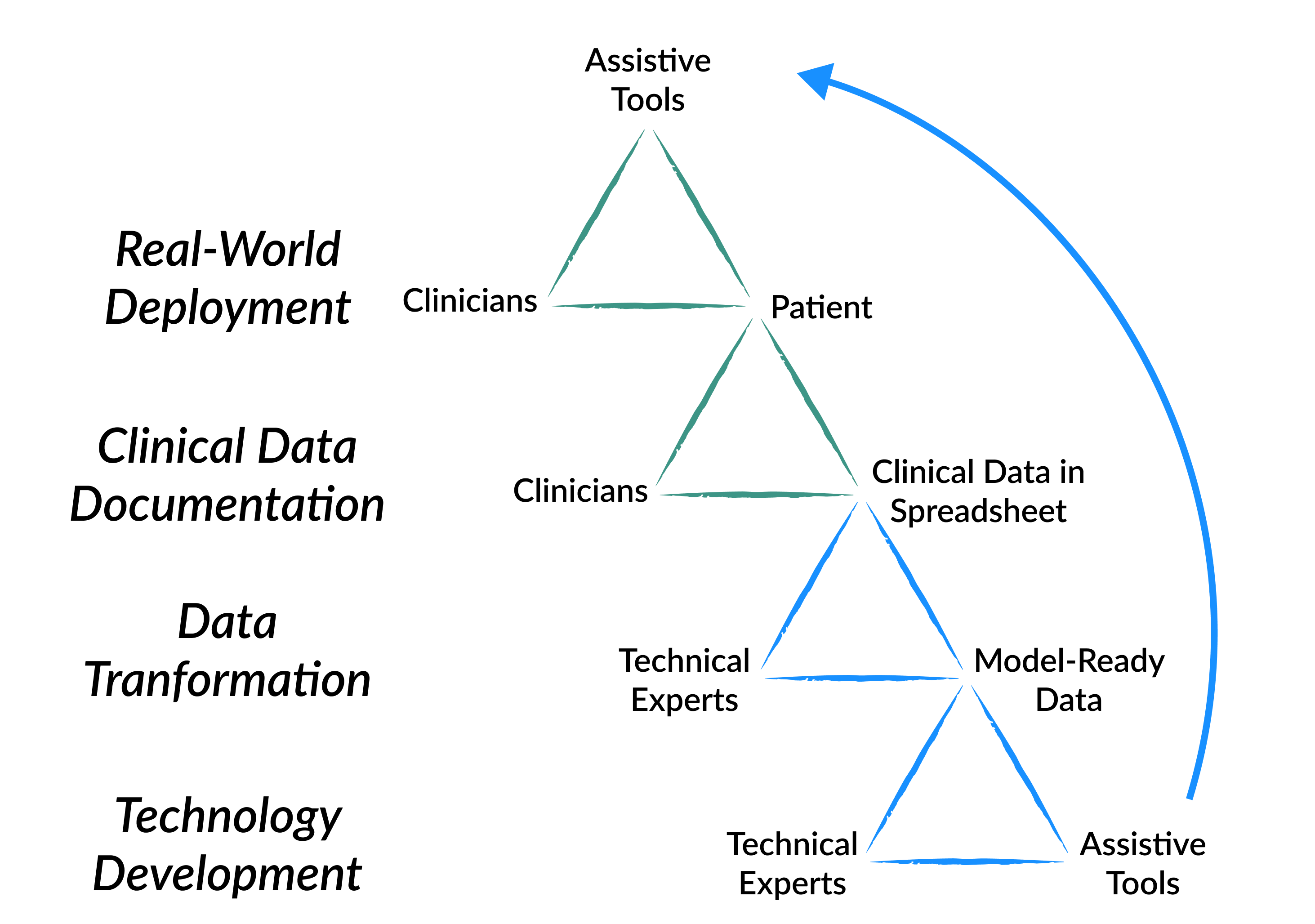}
    \caption{\add{Activity triangle for technical flow in blue. The green color represents the clinical workflow that generated the clinical data for the technical team.}}
    \label{fig:activity-network-clinicalNLP}
\end{figure}

\add{While clinicians focus on identifying and capturing clinical nuances for patient care, technical experts emphasize computational tasks such as training and validating machine learning models. The steps related to the data transformation process in technical experts' workflow often include cleaning low-quality entries, removing noise, splitting data into training and testing sets, and verifying that the label distribution aligns with model assumptions (Figure \ref{fig:activity-network-clinicalNLP}).} \add{One technical participant (P9, Case 1) described his process for working with clinical data and developing a clinical AI system: \textit{“Find LLM to transcribe audio data to text data, very clearly split speaker language timestamp, let clinicians use machine learning to transcribe and have UI interface for clinicians to do checking manually and easily.”} During this process, P9 explored different audio LLMs and commercial products such as "Microsoft, OpenAI Whisper, iFLYTEK" and clearly identified separate clinical vs. technical goals for this clinical AI project:} 

\begin{quote}
\textit{\add{"For audio, if the clinician want to find patterns, the clinician needs to listen to the video second by second and do all the annotations manually. For us, we find out a way to improve efficiency of their work and reduce their time spent on audio annotation. Our goal is to let them don’t spend too much time, and find out exactly what they need to see from the audio."}} \end{quote}

 \add{In this case, the \textit{motive} of technical experts shifts to interpreting and transforming the clinical data into the technical taxonomy that aligns with the requirement for applying machine learning models and evaluation pipelines. Additionally, clinical data can remain opaque to those who lack clinical training and knowledge, despite these non-clinical experts being tasked with understanding and using such data to develop computational models. Differing from Clinician P5 who performed collaborative bilingual data annotation with another clinician, P9 independently evaluated multiple AI tools to identify the best model performance:}

\begin{quote}
\textit{\add{"We try to split the language (Chinese/English) then give to LLM to transcribe everything, another method is split speaker first, let LLM to transcribe audio to text, and see which combination method is the best. Then we realize maybe the split the speaker first use PyAnnotate, use OpenAI Whisper to transcribe is the best part."}}\end{quote}

\add{Dealing with different types of multimodal data that clinicians curate via telehealth bring added challenges for AI teams for data processing. For example, another technical participant P11 who also worked with P5 in Case 1 described how the research team developed a video annotation tool that extracts raw audio data (mp3 format) from the clinician video telehealth sessions:}

\begin{quote}
\textit{"When we use the tool we need another JSON format of machine generated transcription that could be acquired using the release code. Using the AI large language models (LLMs), these are the raw machine generated. After we use these tools to generate machine generated data, we generate a CSV/JSON file."}
\end{quote}

\add{From an AT perspective, data serve as a crucial \textit{object} of shared activity but become embedded in distinct professional contexts (i.e., communities, division of labor, and tools). A technical expert acts as a \textit{subject} who adopts computational frameworks (the \textit{tools}) to transform clinical data (\textit{object}) into model inputs.} After establishing a technical data taxonomy and finishing the data transformation based on the taxonomy, technical experts proceed with designing and refining predictive algorithms, such as to classify patient observations or predict clinical decision-making.
Technical experts usually evaluate these algorithms with quantitative metrics such as accuracy, recall, and precision, and provide performance reports back to the clinical team.

\subsubsection{AT Hierarchies for Understanding Interdisciplinary Workflows}

\begin{figure}[b]
    \centering
    \includegraphics[width=0.7\linewidth]{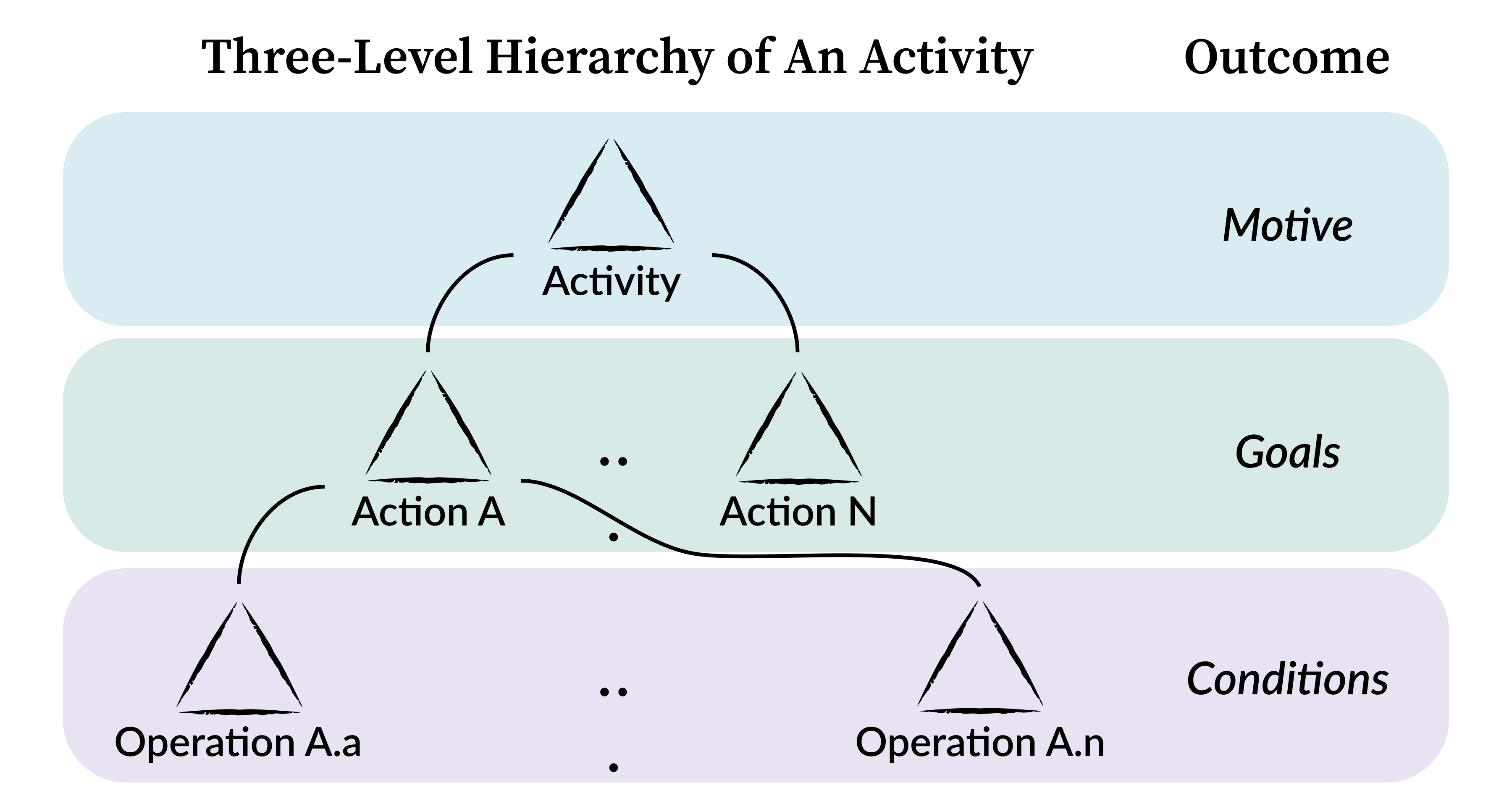}
    \caption{\add{Three-level activity hierarchy showing hierarchical structures activities represented by "actions" which lead to specific "outcomes" and "operations" which lead to specific "conditions."}}
    \label{fig:activity-hierarchy-structure}
\end{figure}

One critical concept of AT lies in the \textbf{hierarchy of an activity}.
Initially proposed by \citet{leontiev1978atividade}, where he stated, ``\textit{We call a process an action if it is subordinated to the representation of the result that must be attained}.'' Later, \citet{kaptelinin2012activity} proposed a three-level AT hierarchy to reflect the internal expansion of one particular activity, as shown in Figure~\ref{fig:activity-hierarchy-structure}.
Specifically, activities are not standalone and disconnected from other activities; instead, activities are hierarchically expansive and constituted by a series of sub-level activities, which are the ``actions''~\cite{leontiev1978atividade}.
When an action becomes an automatic and unconscious behavior, it would be recognized as operational and classified as ``operation'' with respect to AT.
Operations, unlike activities and actions, usually do not have explicit goals. 
In our case, we primarily focus on the activity-action hierarchy and care less about the operations because our focus centers on the processes with clear motivation and objectives in stakeholders' workflows.

\begin{figure}[t]
    \centering
    \includegraphics[width=0.9\linewidth]{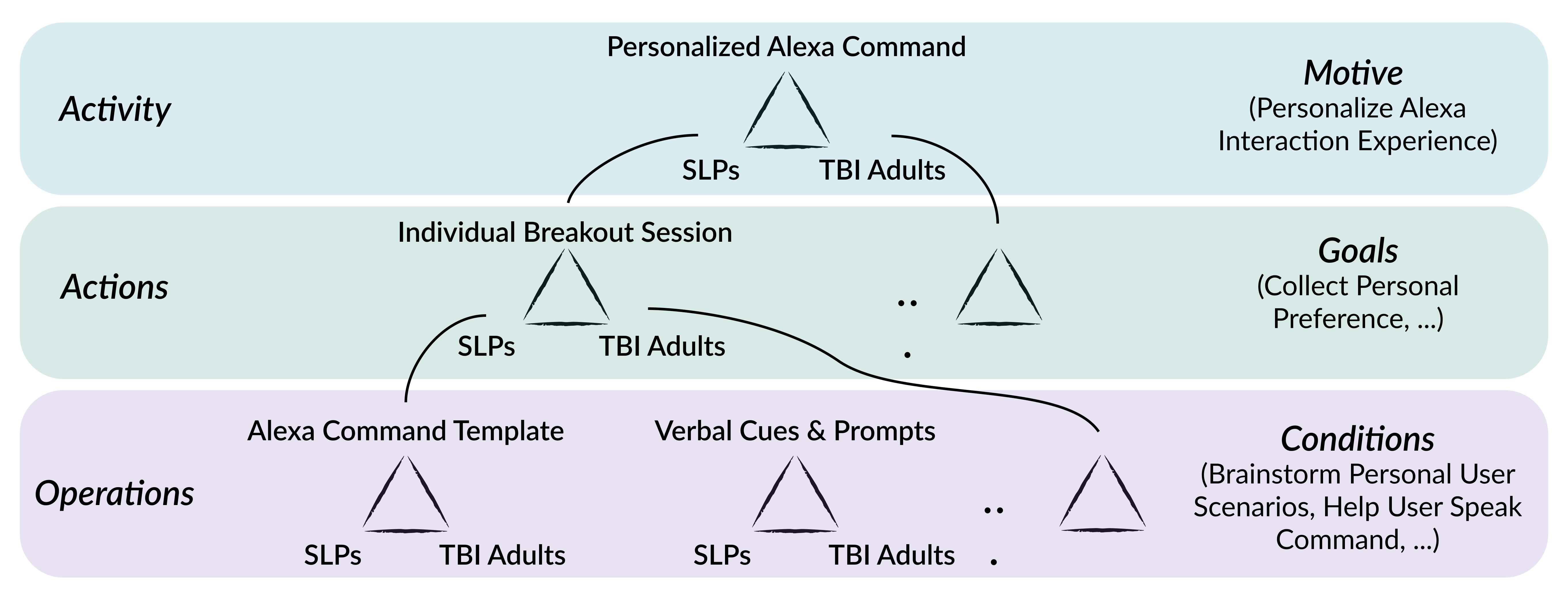}
    \caption{\add{Activity hierarchy of clinical workflows (Case 2), showing activity triangles depicting action and operations as well as corresponding goals and conditions.}}
    \label{fig:activity-structure-clinicalNLP}
\end{figure}

In addition, two activities can be ``chained'' with each other if one component of one activity is shared by a different component of another activity with a direct causal relationship, i.e., ``consumed by or produce for'' the corresponding component of another activity~\cite{spinuzzi2008network}.
In the context of clinical AI team collaboration, we can apply AT's hierarchical framework to the workflow of clinical or technical experts. For each workflow, we can decompose the top-level activity into an activity hierarchy with a series of ``chained'' action-level processes. For instance, clinicians engage in a high-level activity of patient care, which is composed of a series of ``chained'' clinical actions, including source data collection, data taxonomy creation, data transformation, and decision-making \add{(Figure~\ref{fig:activity-structure-clinicalNLP})}.

\begin{figure}[h]
    \centering
    \includegraphics[width=\linewidth]{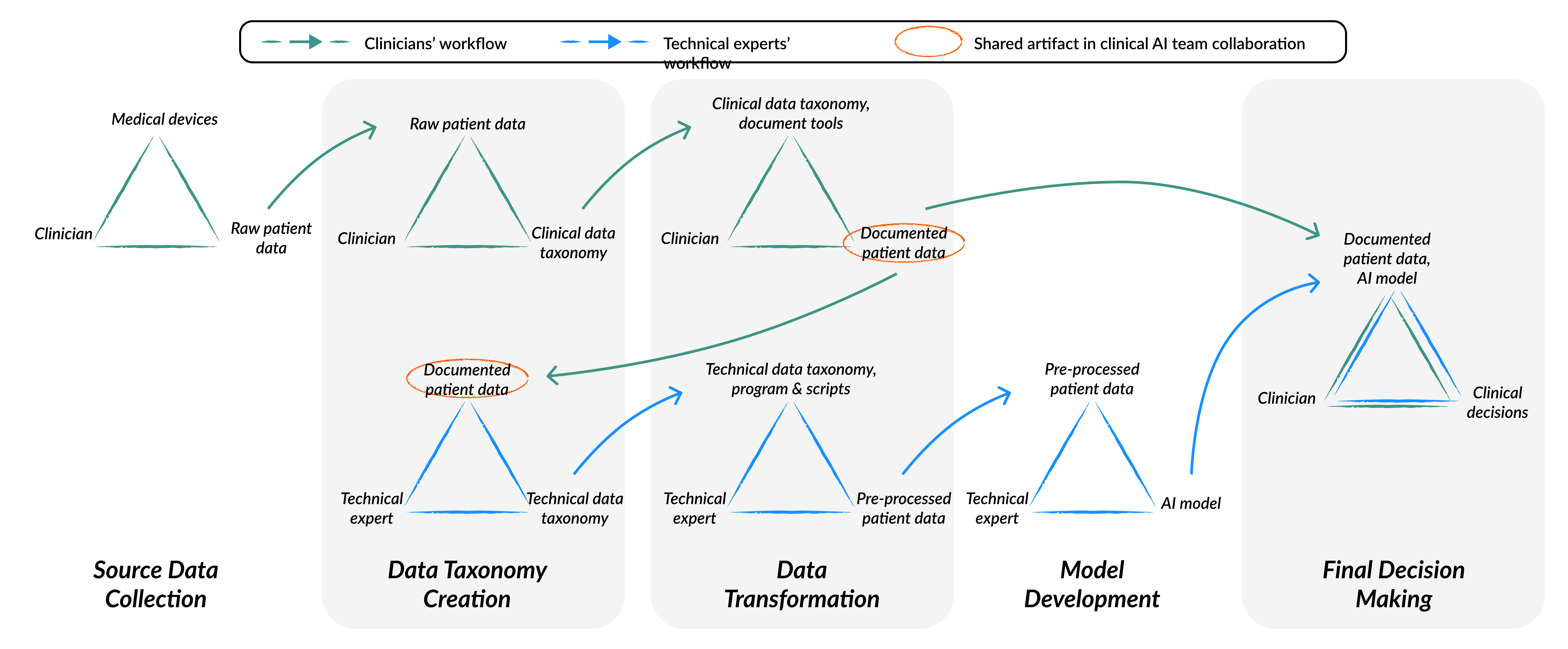}
    \caption{AT network of the clinical AI team collaboration with clinical and technical experts. \add{Common stages are identified as well as shared artifacts "documented patient data" in the distinct workflows of both stakeholders (clinical vs. technical).}}
    \label{fig:activity-network-bothteams}
\end{figure}

By comparing these AT hierarchies of different stakeholders in parallel, we can easily observe the significant overlap between the experts' workflows with respect to the objective of actions. 
\add{Specifically, data taxonomy creation and data transformation are shared across both workflows.
Misalignment becomes visible when one group's sub-activity depends on an artifact the other group has produced in a format ill-suited for downstream use. For instance, we see that clinicians code raw data according to clinical standards, while technical experts refine that data for machine learning pipelines as shown in Figure~\ref{fig:activity-network-bothteams}. 
Although both sides transform the same source data, they do so to achieve different goals via different workflows. 
This difference often leads to tension when classification taxonomies and the definition of ``validity'' in different domains are not mutually understood.} \add{One technical participant (P6, Case 2) explained:}

\begin{quote}
\textit{\add{“The most challenging is in the communication part. I need to understand their data, format their motivation, and they will explain it in the perspective of the clinicians, but I will understand it in the perspective of an NLP expert. So this may have a gap between the clinicians and me.”}}
\end{quote}

While these reports are valuable, clinicians may often struggle to conceptually connect these model performance metrics to the practical realities of patient care.
If clinical experts cannot see how or why model outputs might influence real-world clinical decisions, they may become skeptical about the model's trustworthiness or \re{are} hesitant to adopt it.
After all, our findings underscore how the same artifact (patient data) can follow a second arc of transformation within a different \textit{community} with its own \textit{rules} and \textit{division of labor}.

\subsubsection{Networking Activities Across Disciplines In Team Collaboration}

\begin{figure}[!h]
    \centering
    \includegraphics[width=0.6\linewidth]{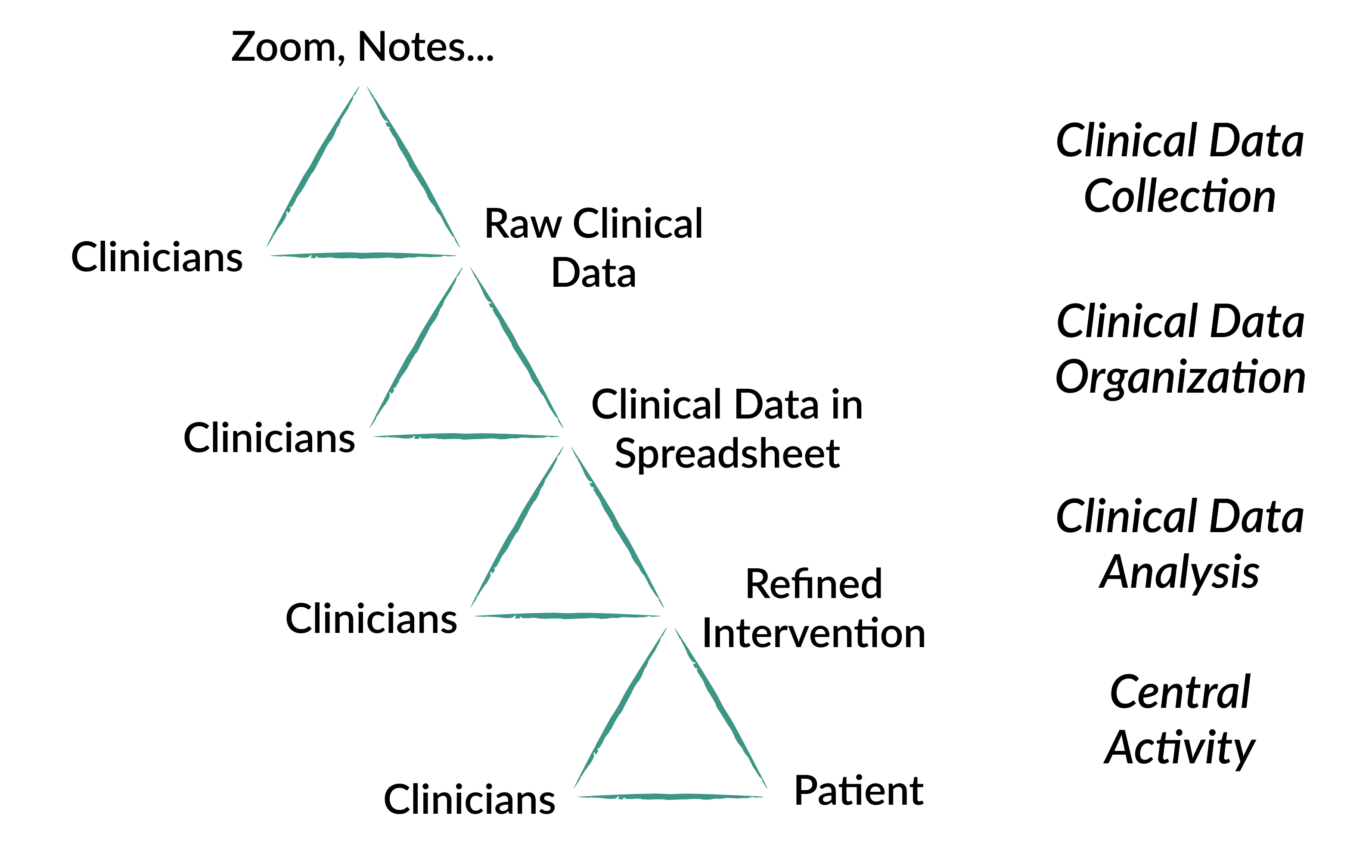}
    \caption{\add{Network of clinical activities illustrating a sequence of data transformation within the clinical team.}}
    \label{fig:activity-network}
\end{figure}

An AT hierarchy is not sufficient to capture how different actions interconnect and what particular challenges were encountered for various workflows. The concept of the \textbf{network of activities}~\cite{engestrom2015learning, bodker2021through} addresses this gap by examining the relation between activities. As defined by \citet{bodker2021through}, ``\textit{... in which one activity’s object could function as another activity’s mediator}.'' The concept of the AT network is critical for HCI research as it supports an in-depth understanding of collaborative work, which is of significant benefit to the human-centered design of assistive computational systems~\cite{nardi1996studying, kaptelinin2009acting}. In clinical AI collaboration,  when clinical experts generate annotated data, they effectively produce an output that the technical team consumes as input. Conversely, once a model has been trained, clinicians might want to interpret its predictions or understand its errors. Through the visual representation of the activity network of clinical AI collaboration shown in \add{Figure~\ref{fig:activity-network}}, it becomes clear how ``clinical data'' is central to both clinical and technical workflows.
This visual representation is very beneficial to help HCI researchers tease out the expert's workflow in team collaboration, particularly the purpose, information needed, tools used, and any related details in each action.

\subsection{Tensions, Challenges, and Coping Strategies}

While both clinicians and technical experts center their work on data, they often face persistent communication barriers caused by widely divergent knowledge boundaries, professional training differences, and domain-specific languages with specialized vocabulary and jargon. \add{When these professional workflows converge during team collaboration, communication difficulties frequently arise because of misaligned taxonomies, divergent expectations about data formats, and the use of different specialized languages. In this subsection, we first unpack different points of tension when interdisciplinary researchers engage in their unique workflows (Section 4.2.1), then identify the knowledge gap that leads to communication challenges (Section 4.2.2), and examine the communication barriers impede collaboration can be addressed via two primary coping strategies: \textbf{using data as a boundary object} (Section 4.2.3) and \textbf{relying on knowledge brokers} (Section 4.2.4). When interdisciplinary researchers become knowledge brokers who can effectively utilize boundary object as a mediating artifact, they can successfully address the communication barriers and collaboration challenges via a plethora of coping strategies.}

\subsubsection{Points of Tension in Convergent Workflows}

Although clinicians and technical experts rely on systematic approaches to code and label clinical data from clinicians' observations, they create and define divergent coding themes and labeling categories based on their disparate domain-specific expertise.
The clinical coding themes of the source data may not map directly onto the technical taxonomy because the clinically meaningful categories may not correspond to an algorithm's data schema. \add{As technical participant P7 (Case 1) noticed that although following the same codebook, clinicians'
data collection and analysis inherently easy to organize into a uniformed computational language. She stated that: \textit{"I found in the spreadsheet is that..it really comes from different annotation styles, annotation habits, different combinations...this really gave me a misunderstanding or misconception...how the alignment happened in those areas? And I think if the clinician agreed upon some like unified styles, or if in other cases, the the categories of different behaviors can be labeled numerically"} then her NLP process would be a lot more easier.}

Similarly, a subtle clinical distinction may be overlooked by a broadly defined technical label. \re{For example, the word "model" for clinicians in the context of speech-language therapy refers to "providing a clinician-led demonstration to show patients how to accurately produce speech or language." Furthermore, in Case 2 (VAT project), the model specifically refers to "the demonstration of how to say an Alexa command using the correct discourse." However, for technical researchers, "model" means "machine learning systems or algorithms to interpret human speech and language." During collaborative communication such as project meetings, these terms (e.g., model, prompt) are constantly used by clinical vs. technical teams, but could refer to entirely different concepts.} These mismatched taxonomies can lead to confusion when technical experts attempt to understand the clinical data, or when clinicians try to understand the model predictions and results. AI expert P12 \add{(Case 1)} observed that when applying automatic classifications using LLMs to clinicians' clinical data, the LLMs may identify novel patterns that can demands insights and interpretations from both clinicians and technical experts:
\begin{quote}
\textit{"I think the hard part is to let them (clinicians) know why parental behavior is important, what type of parental behavior is important, and how to classify them. Differentiating different types of parental behaviors are also very hard for clinicians to clarify, it’s hard for both of us (as clinicians and as AI experts)."}
\end{quote}

Such confusion may accumulate over time if not resolved promptly and significantly impact later project stages, such as model evaluation or system deployment. 
This tension underscores the dual role of clinical data.
The clinical data acts as a central artifact that bridges the professional workflows of two disparate professional domains during team collaboration, yet it could also introduce confusion when collaborators interpret the same data in conflicting ways.
According to the AT framework, collaborators share an overarching \textit{object} (original patient observations).
However, they apply different \textit{tools}, adhere to different \textit{rules}, and operate within distinct \textit{communities} and \textit{divisions of labor}, all of which shape the output differently. 
Data thus function as a \textit{shared artifact} being repeatedly transformed through domain-specific lenses. \add{Applying AT framework and using activity triangles as an illustration, Figure \ref{fig:our-activity-triangle} offers a representation of how tensions in communication can be resulted from multiple core conflicts related to several key components of AT (e.g., "division of labor" due to unclear task responsibility and handoff, misalignment in "domain-specific rules" due to divergent professional practices by clinical vs. AI experts). By considering the \textit{division of labor}, designers should also consider when AI should engage in communication, what information should be presented, and which type of interactions should be leveraged so that AI systems do not interrupt the collaborative workflow but rather engage at necessary stages. This vision requires careful interface design to ensure that AI recommendations or explanations do not introduce new forms of misunderstanding. AT clarifies that simply adding computational functions is insufficient if those tools ignore the broader social and institutional contexts in which collaborations take place.}

Further, the shared artifact in AT networks can expose tensions rooted in knowledge gaps, terminological inconsistencies, or unaligned goals, for instance, where the team needs to align their understanding to a common ground.
These tensions underscore how two differently motivated activities sometimes rely on a single mediating artifact but interpret it in divergent ways.
If the data is coded in a purely clinical taxonomy but never translated into computational labels, the technical team cannot effectively use the data to develop the model. However, incongruent rules, norms, and divisions of labor can introduce contradictions. 
For instance, the institutional policy that restricts how clinicians label patient data for legal protection might conflict with the flexible labeling approaches needed for effective AI development.
Our data suggest that the \textit{division of labor} (e.g., who is responsible for data cleaning, how final decisions are validated) is especially critical for mitigating the aforementioned tensions. 
Clarity on each team member's responsibilities, combined with explicit agreement on which representations of the data matter for each task, can potentially mitigate some of the misunderstandings.

\subsubsection{Knowledge Gaps and Domain-Specific Languages}

The whole workflow of technical experts in team collaboration builds on the prerequisite of understanding clinical data to support technical decisions for model development.
However, understanding clinical data is a non-trivial task for non-medical experts because the data are documented by following the clinical taxonomy infused with professional knowledge.
The lack of clinical experience and knowledge often impedes technical experts' ability to understand clinical data and taxonomy.
Clinical experts, similarly, do not possess the technical knowledge to understand the model-predicted data and technical results presented by the technical experts.
\re{As a result, clinical experts may have difficulty determining whether model outputs would be useful in their workflow or trustworthy enough to inform practice.}
These difficulties often stem from specialized languages spoken by clinical and technical experts. 
The specialized language of each discipline is rooted in the beginning of specialized education and the ongoing practice of domain experts. 
Such a language includes special jargon, concepts with domain-specific meanings, and abbreviations that only people who have gone through the same education and training are able to use and understand proficiently.
Over the years of practice and experience, domain experts have become increasingly familiar with using domain-specific vocabulary while talking about professional content with other people from the same background, in other words, those who speak the same language.

For instance, clinical experts leverage a professional vocabulary to describe and document their clinical activities, such as observations, diagnoses, and interventions. 
Such vocabulary is learned and mastered by all clinical professionals with the same specialty during their medical education and training, which allows clinical experts to communicate with each other without ambiguity during their daily clinical workflow.
Similarly, technical experts also learned to use professional language with specialized jargon and concepts to describe and understand technical specifications (e.g., data, computational algorithms, and tasks) consistently while engaging in communication with other technical experts.
While domain-specific languages facilitate seamless communication within the discipline, such exclusive languages tend to hamper interactions across different disciplines.
From an AT perspective, the divergent languages being used by different domains, as well as the domain-specific knowledge behind the languages, highlight \textit{rules} and \textit{tools} that reflect the internal norms of each \textit{community}.

In interdisciplinary team collaboration, experts in one domain, no matter whether clinical or technical, may not realize that they are using a specialized vocabulary while explaining their domain-specific knowledge to non-experts.
For instance, when clinical experts fail to accurately understand certain technical terms used by technical experts and seek further explanation, the additional explanations provided by technical experts may inadvertently involve the use of more specialized vocabulary.
In such communication scenarios, it is easy for either side to experience severe frustration and potentially cause a dispute due to the prolonged inability to understand what the other party is explaining or the countless attempts to explain their points, but only to be misunderstood repeatedly.
One clinical participant highlighted the depth of this knowledge gap when interpreting the data:

\begin{quote}
    \textit{``I guess we had a little miscommunication, but I guess that was the only time when we were probably using different kind of terminology. Or, I don't know if it was something to do with, probably the terminology that we didn't understand.''} \ \add{(P2, Case 2)}\
\end{quote}

Another participant further emphasized the importance of reducing technical barriers  to when communicate with clinicians:

\begin{quote}
\textit{"I don’t use any terminologies to talk to clinicians. When conducting a NLP research we need it to be clinically user friendly so we will try to use clinical language instead of NLP language. Most clinicians have the master’s degree and statistical training, all of those terminologies (machine learning, AI agents) are not usable for them in their product, but when we are trying to sell them the product we can show them how it works. It’s not necessary to tell clinicians the terminologies; it sounds overwhelming."} \add{(P12, Case 1)} 
\end{quote}
Unresolved communication issues can stall data handoffs and model evaluations. 
A clinical expert may take extensive time clarifying what an annotation means, while a technical expert may repeatedly run into obstacles trying to re-label or merge data categories due to misunderstandings. 
Participants noted that each re-annotation pass risks introducing new inconsistencies, particularly if not guided by a shared understanding of terms. 
This dynamic aligns with prior CSCW literature on collaborative data work, but here it is amplified by the stakes of clinical practice. 
For example, a misinterpretation of a label might cast doubt on the validity of an entire model’s predictions.


\subsubsection{Using Data as a Boundary Object}

Clinical experts and AI experts would naturally engage in communication for explanation or clarification when a miscommunication caused by knowledge gaps emerges.
However, the different languages they speak would often further exacerbate the difficulties in addressing miscommunication.
To mitigate miscommunication, one particular method was brought up by both experts during the interviews, in which they leveraged clinical data as ``boundary objects.'' 
In CSCW literature, a boundary object refers to a stable identity across multiple communities of practice but allows for different interpretations~\cite{star1989institutional}. 
\add{Our analysis shows how clinical data specifically serves this function, acting as a concrete anchor to ground abstract clinical and technical terminologies.}

When an interdisciplinary team focuses on a specific spreadsheet, video segment, or text chunk, they focus on understanding a tangible artifact of the source information underneath the domain-specific taxonomies. 
In particular, clinical data are transdisciplinary, which conveys the same underlying source information collected from the clinical scenario and allows different domain experts to interpret it differently with their expertise.
In this case, clinical data are ``plastic'' enough to be transformed into various representations (e.g., spreadsheet rows/columns for clinical experts, model-ready formats for AI experts) with different taxonomies, yet remain ``robust'' enough that experts from both disciplines recognize it as referring to the same underlying information (i.e., the patient or clinical event). 
By maintaining this shared reference point, clinical data serve as a reliable alternative to verbal explanations and domain-specific jargon, regardless of the specific format. One AI expert participant noted that by sharing annotated samples and providing them as templates, everyone can align on how future data entries should be structured. In this way, concrete data instantiates an anchor point that mediates between specialized vocabularies: \textit{``I will complete some part of the data first, and I will send them a sample, and they may complete the remain data by following the sample I provided.''} (\add{(P6, Case 2)})

Another AI expert participant commented that the benefit of using the multimodal data from different formats enabled her to regain clinical insights:

\begin{quote}
\textit{"Using both video and audio recordings in parallel to the written transcript has been really helpful to me, especially because the transcribing features that don’t necessarily capture all the information spoken (e.g., people speaking over each other, speaker recognition, defining this voice comes from this person, be able to go back and identify who exactly is being talked about)."} \add{\(P10, Case 2)\)}
\end{quote}

Using clinical data as a boundary object enables collaborators to ``talk through'' actual examples rather than relying solely on abstract explanations.
Clinicians can easily explain why certain patient attributes or symptoms are clinically significant, and technical experts can demonstrate how those attributes might be encoded or processed for downstream AI prediction tasks. 
This methodology allows for more efficient clarification of misunderstandings and a tighter feedback loop when refining project scope or validating model outputs. 
If disagreements arise, the team can use data representations with examples as a demonstration to precisely locate where the gap in understanding occurs and make revisions to seek agreement.

\subsubsection{Coping Strategies by Knowledge Brokers in Interdisciplinary Collaboration}



While clinical data can serve as boundary objects to help anchor interdisciplinary discussions, persistent linguistic or conceptual gaps remain unsolved due to language barriers between specialized domains.
Within the interdisciplinary teams, there could be individuals who have enough familiarity with both domains and act as ``knowledge brokers''~\cite{kimble2010innovation, lomas2007between, pawlowski2004bridging, meyer2010rise, caccamo2023boundary}.
These individuals may practice expertise in the intersection of two domains because they might have learned knowledge about the other domain from their prior interdisciplinary collaboration experience or have gone through expert training in both disciplines.
These people serve as critical communication mediators by spotting misunderstandings and translating specialized knowledge into a format that can be uniformly understandable to both parties.
Our thematic analysis identified three coping strategies that were effectively utilized by knowledge brokers.

\paragraph{(1) Familiarize with Each Domain’s Knowledge Boundaries.}
Knowledge gaps are fundamentally defined by the gaps between the knowledge boundaries of each professional domain.
Competent knowledge brokers need enough grounding in the knowledge boundaries for the involved domains in interdisciplinary team collaboration.
Specifically, the knowledge boundaries encompass two types of information: \textbf{domain-specific knowledge} and \textbf{specialized language}.
In clinical AI collaboration, domain-specific knowledge often encompasses medical or clinical terms (e.g., diagnostic labels, standard interventions) and the fundamentals of machine learning (e.g., data labeling, model evaluation metrics).
For each domain, a specialized language of jargon, concepts, and abbreviations was formulated to define and explain domain-specific knowledge.
Domain experts learned the specialized language uniformly 
and uses such language in their professional workflow.
As a result, a sufficient familiarity with the knowledge boundaries ensures knowledge brokers have sufficient information to identify when and what information will cause misunderstandings in team collaboration. A technical expert captured his process for dealing with ambiguous information out of his domain expertise:
\begin{quote}
\textit{"The model sometimes cannot capture those part, sometimes the model is gonna miss that part. We dont know what’s a good method in clinicians’ ears... In our design, we can't ignore those information because we don’t know if those information matters to clinicians...Keep this information and put everything in “other” ask clinicians for advice, instead of scratching this information out...I do have experience working with different teams, everyone has their own domain knowledge and it’s hard to discuss concerns/goals, and also hard to ask what has been done before.} \add{"\ (P9, Case 1) }\ 
\end{quote}

\paragraph{(2) Identify Miscommunications Promptly}
Given the awareness of knowledge boundaries and specialized vocabularies in both disciplines, knowledge brokers need to be able to identify signs of miscommunication promptly.
For instance, knowledge brokers could identify certain pieces of information stated by one side (e.g., clinicians) that lie outside the knowledge boundaries of the other side (e.g., technical experts).
Miscommunication situations include when some concepts and phrases convert totally different domain-specific information in different disciplines, but experts from both groups are not aware of the difference. In these situations, experts from both groups would falsely believe they correctly understand each other, and the signal for miscommunications could be difficult to detect. \re{For example, P13 (Case 2) and P11 (Case 1) are both technical experts working on different clinical data from different projects. However, both encountered ambiguous annotation/clinical documentation practices that are difficult to articulate. P13 had to explain automatic speech recognition (ASR) and the concept of "how large audio models provide text transcription" to clinicians without technical jargon, so she came up with an example as if she had to teach the concept "to the primary school kids" when communicating with clinicians:}

\begin{quote}
\textit{\re{"If you have a food machine in front of you and you want to get a coke from this machine, this machine is like a large number model. You want to get a coke. And how do you get a coke? You usually need to pay and you need to select the number right? And for large, large language model, it's the same thing. You need to give the input like our audio data, and then you need to tell what exactly you want, and the large language model can just give you the result."\add{(P13, Case 2)} }}
\end{quote}

\begin{figure}[!t]
    \centering
    \includegraphics[width=\linewidth]{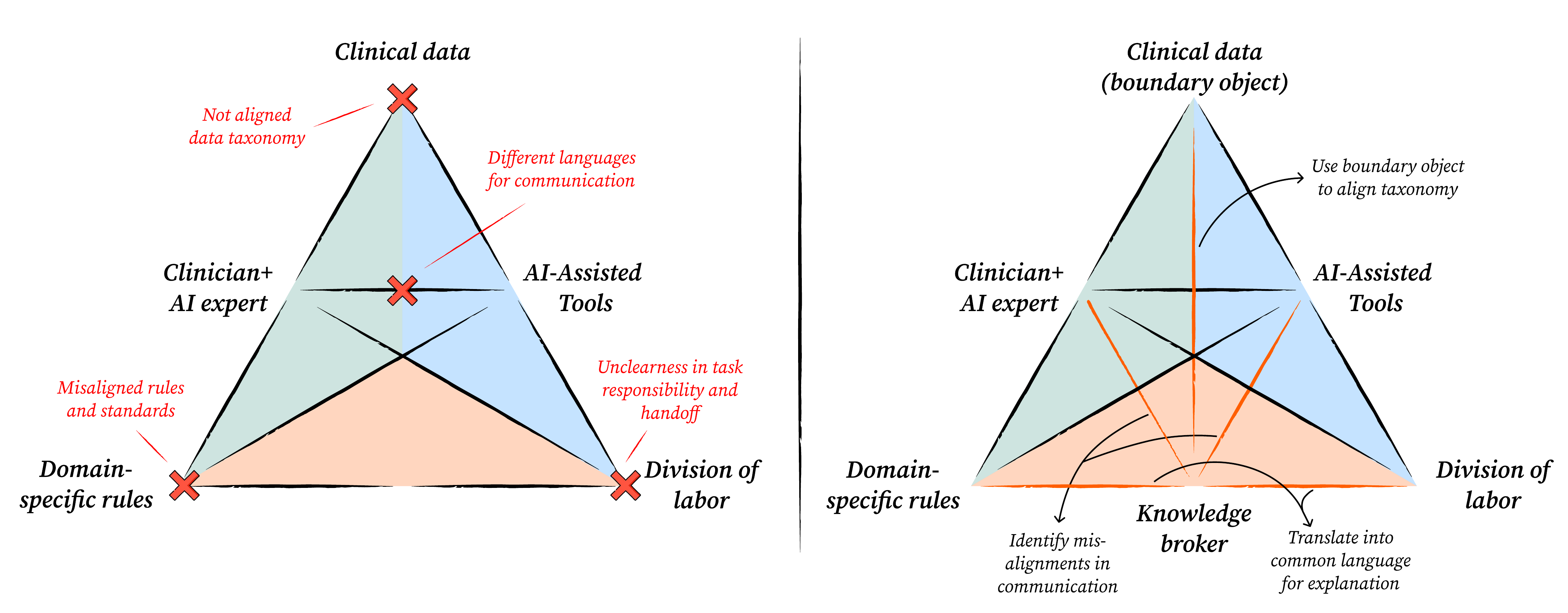}
    \caption{The activity triangle views of interdisciplinary team collaboration in Clinical AI. Different background colors represent the knowledge boundaries of the clinical, technical, and shared domains. The left triangle demonstrates the ineffectiveness between clinical and technical experts \add{when attempting to collaborate on AI-assisted tools without a knowledge broker's assistance, whereas the right triangle demonstrates how knowledge brokers effectively utilize clinical data as boundary objects to facilitate seamless collaboration. }}
    \label{fig:our-activity-triangle}
\end{figure}

\re{P11 performed initial clinical data work by "looking at spreadsheet ... (with) very colorful...red, blue and orange" which was not explained clearly and consistently, resulted in difficulty in sense-making. However, through multiple ongoing research meetings, he gradually learned that work from his side is "to mostly accommodate the things that don't really make sense...at the first place, but they could be certain any like important purposes in to other people like in clinicians." Eventually, he learned specific behavioral classification reasoning through the color coding and commented: "people cannot really get a clue of lean the logic behind why it is so. And I think that's that's a challenge not only for clinicians, but as well as as well as for researchers ourselves as well."} 
The ability to identify miscommunications promptly is critical to facilitate effective communication because misunderstandings can stack up if left unaddressed and lead to more substantial confrontations at a later stage. Another technical expert described on her experience communicating and collaborating with another clinician on a shared transcription task:

\begin{quote}
\textit{"I collaborate for data cleaning with another clinician who has also taken on the burden of going through the data manually, listening to transcripts, comparing/contrasting, and catch inconsistencies. That part of the collaboration - because she comes in a bit later - getting to explain to her all the things/challenges I had noticed along the way, and how I resolved it, learning how to re-communicate instructions that I understand completely but making assumptions that may not be the assumptions on her end, is interesting. Having an open dialogue and collaboration - `oh I ran into this challenge' - go back and patch those challenges."} \add{(P10, Case 2)} 
\end{quote}

\paragraph{(3) Translating Specialized Languages into a Common Language}
Being able to identify knowledge gaps is something that can be easily achievable by people who have expertise or knowledge in both professional domains.
However, being able to address the communication challenges caused by knowledge gaps is a much more complicated task that requires knowledge brokers to carefully decide what approach they would use to address the issue.
Specifically, such a decision often leads to the question of what language knowledge brokers need to use that will not introduce new confusion.
Importantly, knowledge brokers must translate domain-specific jargon into a specific language that both sides can uniformly understand. A technical expert who led the AI work for one project commented on her experience when attempting to translate different goals and communication barriers when collaborating with clinicians:

\begin{quote}
\textit{"We have very different ways of thinking, in the very beginning I tried to understand what’s the problem there I learned later it’s hard for them to identify the problem in the very beginning. I think my personal experience in this collaboration is, we put some efforts in identifying things that AI teams can do in a project, this takes time to finalize. That’s an interesting observation that I have. Other than that, every time we have issues they try to answer us."} \(\add{(P13, Case 2)}\) \end{quote}

\add{In practice, these three capabilities reflect the function of knowledge brokers similar to that of bilingual translators. 
They must continuously switch between clinical and technical frames of reference and ensure no vital details are lost. 
Knowledge brokers are expected to translate different specialized languages into a common language that is used and shared by everyone on the team.
By doing so, different domain experts will be able to understand everything on the same page using the same language.
For disciplines that have very little overlap in the language they use, such as the clinical domain and the technical domain, the common language that everyone is able to understand is often a lay language that does not include any specialized jargon or concepts.
As a result, the knowledge broker needs to translate the specialized language of the speakers into the lay language that everyone can correctly and uniformly understand, and only use the lay language while providing further explanations or clarification during the communication.}

\begin{figure}[!tp]
    \centering
    \includegraphics[width=.85\linewidth]{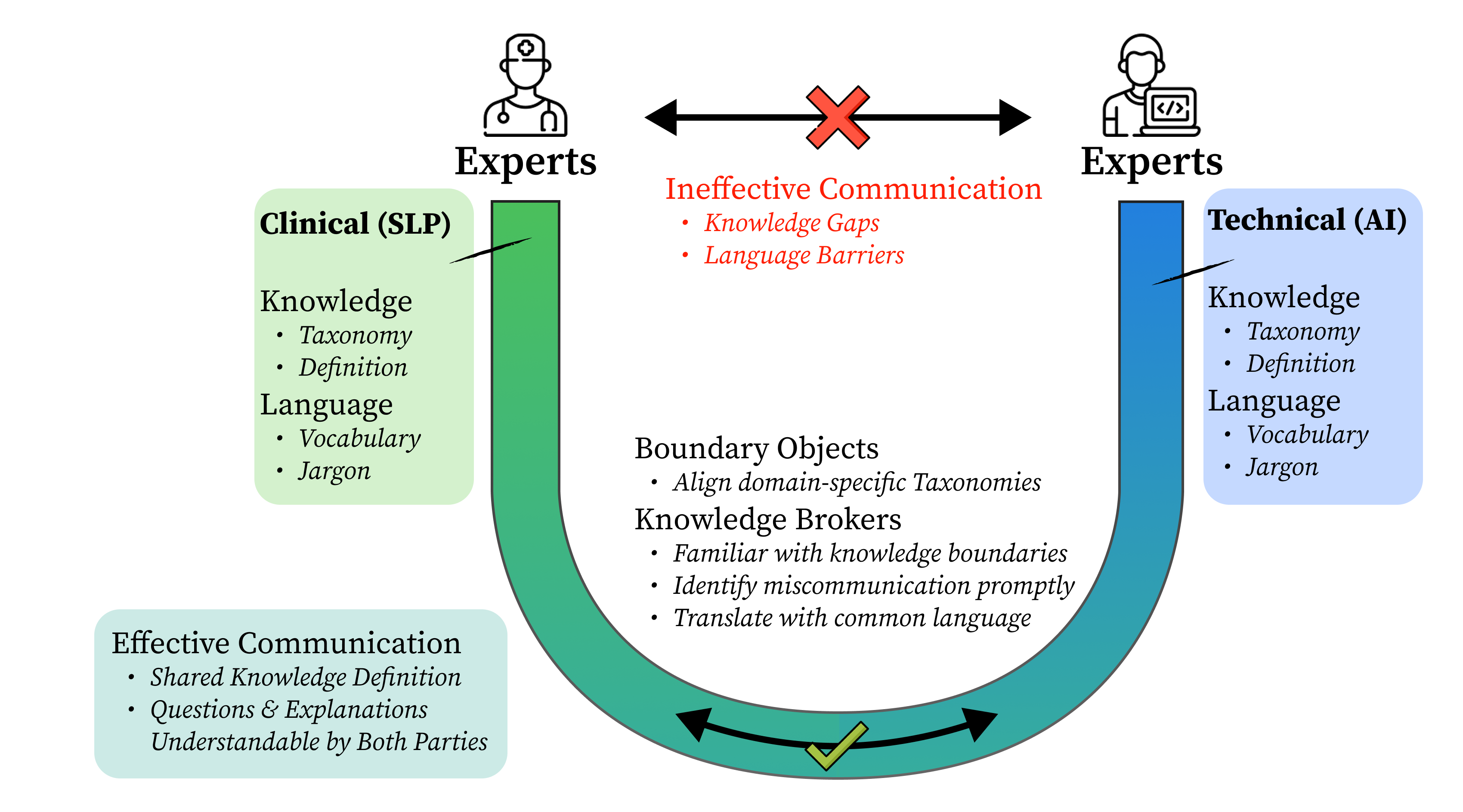}
    \caption{Illustration of communication barriers and coping strategies in clinical AI team collaboration. 
    Direct communication between domain experts is ineffective due to knowledge gaps and language barriers, whereas available boundary objects and knowledge brokers could help bridge miscommunication.}
    \label{fig:knowledge-breaking}
\end{figure}

\add{In addition to these themes, using AT can also shed light on practical design implications for AI assistance in team collaboration with respect to the activities conducted by human knowledge brokers. One coping strategy to bridge the knowledge gaps is to leverage data examples as boundary objects to ``speak'' for the information they represent. The shared object of clinical data traverses multiple communities with different rules, tools, and divisions of labor; still, the language barriers are unsolved. Another strategy is to rely on individuals who are capable of playing the role of knowledge brokers. As shown on the right diagram of Figure~\ref{fig:our-activity-triangle}, knowledge brokers enable the use of a common language between different domain experts to bridge communication difficulties.The suite of knowledge-breaking requirements we identified, which was visually demonstrated in Figure~\ref{fig:knowledge-breaking}, indeed sheds light on important research opportunities in HCI and CSCW to explore how we can leverage advanced AI technologies to better support interdisciplinary collaboration. AI technologies such as LLMs have the potential to achieve the aforementioned prerequisites through carefully designed, human-centered interactions and interfaces. }






In summary, our findings show how clinical data, while central to interdisciplinary AI collaboration, can still trigger deep confusion when clinical and technical experts bring desperate domain-specific perspectives about how to label, process, and interpret data. Communication barriers often arise from gaps between professional knowledge boundaries and divergence in specialized vocabularies.
Two primary strategies--leveraging data as a boundary object and relying on knowledge brokers--can help alleviate these problems. 
Activity Theory further enriches our understanding by situating these challenges in a structured view of each group’s motives, actions, and communal rules. 
This perspective reveals multiple design opportunities for AI systems aimed at bridging communication gaps. 
By drawing on knowledge brokers’ skill sets and anchoring interactions around well-defined boundary objects, emerging technologies such as LLMs can be harnessed more effectively to support truly collaborative clinical AI development.

\section{Discussion}
\label{sec:discussion}

Our study provides a systematic examination of interdisciplinary team collaboration in clinical AI by focusing on how clinicians and technical experts jointly leverage shared clinical data, and manage mismatched taxonomies, specialized terminologies, and knowledge gaps when working on two projects that integrate AI in speech-language pathology assessment and intervention services.
By employing Activity Theory (AT)~\cite{engestrom2000activity, nardi1996context} as an analytical lens to evaluate two case studies, our work \add{investigated beyond surface-level communication challenges to analyze the underlying structural misalignments and conflicts related to professional rules, division of labor, and community norms.}
Below, we reflect on these findings in light of existing CSCW/HCI research by distilling \add{contributions to existing theories}, implications for best practices, design considerations for AI-based knowledge-breaking tools, and opportunities for future work that extend beyond the immediate context of clinical AI collaboration.

\subsection{Leveraging AT \add{to Reveal the Structural Tensions} of Clinical AI Collaboration}

Our findings underscore the critical but sometimes problematic role of data in bridging clinical and technical experts in Clinical AI collaboration. Previous work has highlighted how data and representational artifacts can mediate cross-boundary collaboration and classification work~\cite{star1989institutional, bowker1999sorting, carlile2002pragmatic, carlile2004transferring, mao2019data, pine2022investigating, blackwell2009radical}.
However, we recognized that data remains a ``double-edged sword'' when divergent objectives and specialized data themes shape how clinical and technical experts annotate and interpret the same raw observations. 
AT helps explain why these tensions recur: each domain follows distinct rules and norms while pursuing unique goals, yet both rely on overlapping artifacts and tools \cite{schmidt1992taking, star1994steps}. 
In contrast to scattered, case-by-case analyses of communication breakdowns in healthcare technology collaborations~\cite{chen2020trends, linhares2022clinicalpath, bossen2019data}, our analysis with the AT framework reveals repeated structural misalignment and conflicts across two distinct clinical AI collaborations. 
As AI expands into specialized healthcare domains~\cite{yang2024talk2care, jiang2023health, wu2024cardioai}, a system-level perspective becomes necessary to support more seamless collaborations and practical outcomes.
AT highlights that mismatches in data coding schemes and domain-specific languages are patterned consequences of intersecting activity systems rather than isolated anomalies~\cite{schmidt1992taking, carlile2004transferring}.



\subsection{\add{Mediating Asymmetries with Boundary Objects}}

A key insight from our study is that spreadsheets, annotated transcripts, and other tangible artifacts can function as ``boundary objects''~\cite{star1989institutional, carlile2002pragmatic}, which are flexible enough to adapt to each group's perspective yet stable enough to carry shared information across domains. 
Rather than relying on abstract definitions, interdisciplinary teams can ``talk through'' data examples to establish the aforementioned common ground of knowledge.
This resonates with earlier CSCW work on how physical or digital artifacts can anchor collaborations and clarify misunderstandings~\cite{bannon1997constructing, bossen2002parameters, dourish2004we, hou2017hacking}. 
\add{Beyond simply sharing information, these objects also play a role in mediating the epistemic asymmetries inherent in these teams~\cite{suchman1987plans, carlile2004transferring}.
Such asymmetry is rooted not only in different knowledge but also in control over critical resources, where clinical experts possess the clinical data resources and technical experts hold expertise in computational methods. The mutual dependency between them creates a delicate power balance that must be continuously negotiated.
For instance, a prototype allows a clinician to critique an AI system based on its workflow integration, providing a tangible way to assert their clinical expertise and balancing the power dynamic that might otherwise favor purely technical metrics~\cite{amershi2019guidelines}.
Boundary objects, in this case, may function as sites for situated action~\cite{suchman1987plans}, where the information being carries is not fixed but is actively negotiated as collaborators interact with and through the artifact.} 
In some cases, non-data artifacts such as journey maps or interface mock-ups surface workflow mismatches more directly than raw data~\cite{bannon1997constructing}. 
Nonetheless, boundary objects alone rarely suffice when specialists lack a conceptual bridge to interpret one another's domain-specific terminologies, data themes, and analytic frameworks~\cite{schmidt1992taking, carlile2004transferring, tell2016managingknowledge}.


\subsection{\add{Knowledge Brokers and the Division of Labor}}

Each domain operates with its own specialized terminology, which, though functional within the domain, can create confusion when used in interdisciplinary contexts and reinforce knowledge boundaries~\cite{friman2010understanding, orlikowski1994technological, bechky2003sharing}. 
Our analysis affirms that ``knowledge brokers''~\cite{lomas2007between, kimble2010innovation, caccamo2023boundary} play an indispensable role in interdisciplinary teams by primarily functioning like ``bilingual translators''~\cite{wenger1999communities, meyer2010rise}. 
Their capacity to understand knowledge boundaries, detect terminological differences promptly, and translate specialized languages into a shared working language aligns with earlier studies of organizational intermediaries~\cite{lomas2007between, caccamo2023boundary}. 
\add{This role is often visualized through articulation work~\cite{schmidt1992taking, schmidt1994cooperative}, which describes the often invisible effort required to align interdependent tasks in complex settings. 
From this perspective, the broker is a primary agent of articulation who manages conceptual, technical, and temporal interdependencies across activity systems~\cite{schmidt1994cooperative}.}

\add{Cross-boundary coordination can be understood as operating in a ``trading zone''~\cite{galison1997image, collins2007trading}, a conceptual space where communities develop local contact languages or pidgins to collaborate. 
In our cases, broker-curated artifacts such as data glossaries functioned as the pidgin that enabled productive exchange.}
Expertise is dynamic and uneven across individuals, which requires brokers to continuously assess and adapt their translations. 
In our study, knowledge brokers enabled more productive negotiation of how clinically coded data would be reused for AI modeling and how AI results would be fed back to clinicians in clinically interpretable ways. 
However, as others have noted~\cite{meyer2010rise, caccamo2023boundary}, these individuals are often in extremely short supply, which could magnify the risk of bottlenecks or disruptions as the need for interdisciplinary team collaboration rockets but the availability of knowledge brokers stagnates.

Although boundary objects and knowledge brokers appeared to be central in bridging collaboration gaps, our work confirms that interdisciplinary teams also operate under explicit regulations and institutional guidelines, which are conceptualized as \textit{rules} in AT~\cite{engestrom2018expertise}. 
For clinical AI collaborations, these rules may include ethical considerations (e.g., patient confidentiality under HIPAA~\cite{assistance2003summary}), professional standards for diagnosing disorders, or data-sharing guidelines that affect model development timelines and resource allocation. 
Neglecting such rules can lead to stalled collaborations, even if boundary objects and brokers are otherwise effective. 
Similarly, ``division of labor''~\cite{leontiev1978atividade, spinuzzi2008network} is critical but often neglected by the teams. 
Researchers have long noted that poorly defined responsibilities for different participants can generate friction and mistrust in high-stakes settings like healthcare~\cite{agapie2024conducting, bednarik2022integration}. 
Our findings extend this point by highlighting how divisions of labor \add{and the articulation work needed to manage them} shift at different stages of AI system development.

\subsection{\re{Implications of Best-Practices for Interdisciplinary Team Collaboration}}

\add{One best practice for future teams is to establish shared learning opportunities early in the collaboration to make sure different groups can establish a common ground of domain-specific knowledge. 
Regular knowledge exchange sessions, where clinical and technical experts offer domain-specific training or workshops, can help build foundational cross-disciplinary understanding.
This practice is consistent with previous studies that emphasize the importance of creating shared knowledge spaces to facilitate understanding between fields~\cite{clark1991grounding, olson2000distance, klein2010taxonomy}.
Treating common-ground work as an explicit project activity makes this labor visible and valued rather than ad-hoc~\cite{schmidt1992taking}.
We also recommend co-creating a living glossary to reduce confusion and align definitions from the start; this strategy is consistent with findings on terminology variation and interdisciplinary coordination~\cite{furnas1987vocabulary, bechky2003sharing}. Such an artifact serves to make articulation work visible and distributed across the team, reducing the burden on a single knowledge broker.}

\add{While our cases focused specifically on clinical and technical experts, clinical AI collaborations frequently involve HCI specialists and other stakeholders (e.g., product managers, data analysts, etc.) who bring their own domain-specific perspectives to the teams~\cite{rundo2020recent}. 
Future teams should consider how integrating these additional roles might introduce new terminologies or require new types of boundary objects, highlighting the need for explicit communication guidelines to manage this increased complexity~\cite{clark1991grounding, olson2000distance}.}

\section{Limitation and Future Work}

Although our study offers novel insights into interdisciplinary collaborations for clinical AI, particularly in the field of speech-language pathology, several limitations should be noted. First, our \rev{empirical context is situated specifically within the clinical practice of speech-language pathology and telehealth} applications involving small teams of clinicians and AI experts. The activity triangles, networks, and boundary objects that shaped our analysis reflect this specialized domain and may not map directly onto other domains or team collaboration \add{for broader implications for clinical AI work}. Whether the challenges and mitigation strategies we identified can be generalized to broader or more complex interdisciplinary settings--such as large industrial research collaborations--remains an open question. Second, while we investigated how different domain experts conceptualize, document, and interpret data, we did not delve deeply into the development of interactive systems or interfaces stemming from the collaborative process, such as a web-based assessment platform or an app-based patient-management tool. 
\add{Third, the current case studies consist of collaborations at an early development stage where models are not yet applied in clinical practice. The current analysis emphasizes "data work" such as data input and preparation (e.g., clinicians provide data from patient care, technical experts process data for model pipelines). Therefore, discussions of model output and its clinical utility after technical experts' development are limited.} 

\rev{Our findings point to specific areas for future investigation that build directly on these aforementioned limitations. First, while clinical data served as an effective boundary object in these settings, future research must examine whether these collaborative dynamics and the role of knowledge brokers operate similarly across other highly specialized medical domains for more generalized research contributions to similar clinical-AI collaborative work. Second, although our study did not extensively investigate the design and development of collaborative systems to better facilitate interdisciplinary team collaboration, the analytical perspectives of AT are crucial for more extensive future research on concrete interface design and system prototype development. With the empirical insights on knowledge boundaries and communication needs in clinical and AI workers, future work can make AI-driven solutions more user-friendly and clinically relevant. Additionally, while our participants relied on static artifacts like spreadsheets and transcripts, future work could explore how dynamic or AI-mediated boundary objects might proactively identify and clarify terminological mismatches as they occur, further scaffolding the interdisciplinary data work identified in our cases. Finally, because our analysis primarily captures upstream data work at an early developmental stage, longitudinal studies offer a necessary next step. Tracking how boundary objects adapt over time as models are deployed and clinical teams evaluate downstream AI outputs will reveal their long-term impact on team performance and project outcomes~\cite{bjorn2009boundary, pine2022addressing}. }

\section{Conclusion}

This study examined how clinical and technical experts collaborate on clinical AI projects. 
By adopting Activity Theory (AT) as a structured lens, we \re{found that clinicians and technical experts engaged in distinct yet interdependent forms of data work, which created recurring coordination tensions and communication barriers around domain knowledge, specialized terminology, and data taxonomies. } 
\re{While shared clinical data served as an important coordination artifact, our findings show that data alone could not bridge deeper differences in interpretation and workflow expectations.}
Instead, human knowledge brokers \re{who could move across clinical and technical domains played an important role in translating concepts, clarifying assumptions, and supporting collaboration. }
\re{Overall, this study contributes to CSCW research on interdisciplinary data work by showing how shared artifacts, broker roles, and coordination tensions shape early-stage clinical AI collaboration. 
Future work can examine how these dynamics evolve across later stages of development and whether new tools can better support interdisciplinary coordination in specialized clinical settings.}


\bibliographystyle{ACM-Reference-Format}
\bibliography{Reference}

@article{yamanie2023impact,
  title={The impact of knowledge brokering in health sector and the challenges: A review of literature},
  author={Yamanie, N and Amanda, NF and Felistia, Y},
  journal={Journal of public health research},
  volume={12},
  number={2},
  pages={22799036231167833},
  year={2023},
  publisher={SAGE Publications Sage UK: London, England}
}

@article{gaid2023barriers,
  title={Barriers and facilitators to knowledge brokering activities: perspectives from knowledge brokers working in Canadian rehabilitation settings},
  author={Gaid, Dina and Ahmed, Sara and Thomas, Aliki and Bussi{\`e}res, Andr{\'e}},
  journal={Journal of Continuing Education in the Health Professions},
  volume={43},
  number={2},
  pages={87--95},
  year={2023},
  publisher={LWW}
}

@article{mickan2022using,
  title={Using knowledge brokering activities to promote allied health clinicians’ engagement in research: a qualitative exploration},
  author={Mickan, S and Wenke, Rachel and Weir, Kelly and Bialocerkowski, Andrea and Noble, Christy},
  journal={BMJ open},
  volume={12},
  number={4},
  pages={e060456},
  year={2022},
  publisher={British Medical Journal Publishing Group}
}

@article{douglas2022knowledge,
  title={Knowledge brokering in communication sciences and disorders},
  author={Douglas, Natalie and Oshita, Jennifer and Schliep, Megan and Feuerstein, Julie},
  journal={Perspectives of the ASHA Special Interest Groups},
  volume={7},
  number={3},
  pages={663--668},
  year={2022},
  publisher={American Speech-Language-Hearing Association}
}

@article{cross2023roles,
  title={Roles and effectiveness of knowledge brokers for translating clinical practice guidelines in health-related settings: a systematic review},
  author={Cross, Amanda J and Haines, Terry P and Ooi, Choon Ean and La Caze, Adam and Karavesovska, Sara and Lee, Eu Jin and Siu, Samuel and Sareen, Sagar and Jones, Carlos and Steeper, Michelle and others},
  journal={BMJ Quality \& Safety},
  volume={32},
  number={5},
  pages={286--295},
  year={2023},
  publisher={BMJ Publishing Group Ltd}
}

@article{hashim2007activity,
  title={Activity theory: A framework for qualitative analysis},
  author={Hashim, Nor Hazlina and Jones, Michael},
  year={2007},
  publisher={University of Wollongong}
}

@article{chigbu2019visually,
  title={Visually hypothesising in scientific paper writing: Confirming and refuting qualitative research hypotheses using diagrams},
  author={Chigbu, Uchendu Eugene},
  journal={Publications},
  volume={7},
  number={1},
  pages={22},
  year={2019},
  publisher={MDPI}
}

@inproceedings{crilly2006using,
  title={Using research diagrams for member validation in qualitative research},
  author={Crilly, Nathan and Clarkson, P John and Blackwell, Alan F},
  booktitle={International Conference on Theory and Application of Diagrams},
  pages={258--262},
  year={2006},
  organization={Springer}
}

@article{fereday2006demonstrating,
  title={Demonstrating rigor using thematic analysis: A hybrid approach of inductive and deductive coding and theme development},
  author={Fereday, Jennifer and Muir-Cochrane, Eimear},
  journal={International journal of qualitative methods},
  volume={5},
  number={1},
  pages={80--92},
  year={2006},
  publisher={SAGE Publications Sage CA: Los Angeles, CA}
}

@book{creswell2017designing,
  title={Designing and conducting mixed methods research},
  author={Creswell, John W and Clark, Vicki L Plano},
  year={2017},
  publisher={Sage publications}
}

@article{wu2019large,
  title={Large teams develop and small teams disrupt science and technology},
  author={Wu, Lingfei and Wang, Dashun and Evans, James A},
  journal={Nature},
  volume={566},
  number={7744},
  pages={378--382},
  year={2019},
  publisher={Nature Publishing Group UK London}
}

@article{guhan2025developer,
  title={Developer Insights into Designing AI-Based Computer Perception Tools},
  author={Guhan, Maya and Hurley, Meghan E and Storch, Eric A and Herrington, John and Zampella, Casey and Parish-Morris, Julia and L{\'a}zaro-Mu{\~n}oz, Gabriel and Kostick-Quenet, Kristin},
  journal={arXiv preprint arXiv:2508.21733},
  year={2025}
}

@inproceedings{shamszare2023clinicians,
  title={Clinicians’ perceptions of artificial intelligence: focus on workload, risk, trust, clinical decision making, and clinical integration},
  author={Shamszare, Hamid and Choudhury, Avishek},
  booktitle={Healthcare},
  volume={11},
  number={16},
  pages={2308},
  year={2023},
  organization={MDPI}
}

@article{li2025nerve,
  title={Nerve-Inspired Optical Waveguide Stretchable Sensor Fusing Wireless Transmission and AI Enabling Smart Tele-Healthcare},
  author={Li, Tianliang and Wang, Qian'ao and Cao, Zichun and Zhu, Jianglin and Wang, Nian and Li, Run and Meng, Wei and Liu, Quan and Yu, Shifan and Liao, Xinqin and others},
  journal={Advanced Science},
  volume={12},
  number={4},
  pages={2410395},
  year={2025},
  publisher={Wiley Online Library}
}

@article{osong2025development,
  title={Development of clinical decision support for patients older than 65 years with fall-related TBI using artificial intelligence modeling},
  author={Osong, Biche and Sribnick, Eric and Groner, Jonathan and Stanley, Rachel and Schulz, Lauren and Lu, Bo and Cook, Lawrence and Xiang, Henry},
  journal={PloS one},
  volume={20},
  number={2},
  pages={e0316462},
  year={2025},
  publisher={Public Library of Science San Francisco, CA USA}
}

@article{shaik2023remote,
  title={Remote patient monitoring using artificial intelligence: Current state, applications, and challenges},
  author={Shaik, Thanveer and Tao, Xiaohui and Higgins, Niall and Li, Lin and Gururajan, Raj and Zhou, Xujuan and Acharya, U Rajendra},
  journal={Wiley Interdisciplinary Reviews: Data Mining and Knowledge Discovery},
  volume={13},
  number={2},
  pages={e1485},
  year={2023},
  publisher={Wiley Online Library}
}

@article{nardi1996studying,
  title={Studying context: A comparison of activity theory, situated action models, and distributed cognition},
  author={Nardi, Bonnie A},
  journal={Context and consciousness: Activity theory and human-computer interaction},
  volume={69102},
  pages={35--52},
  year={1996}
}

@book{bodker2021through,
  title={Through the interface: A human activity approach to user interface design},
  author={Bodker, Susanne},
  year={2021},
  publisher={CRC Press}
}

@article{clark1991grounding,
  title={Grounding in communication.},
  author={Clark, Herbert H and Brennan, Susan E},
  year={1991},
  publisher={American Psychological Association}
}

@book{wenger1999communities,
  title={Communities of practice: Learning, meaning, and identity},
  author={Wenger, Etienne},
  year={1999},
  publisher={Cambridge university press}
}

@article{olson2000distance,
  title={Distance matters},
  author={Olson, Gary M and Olson, Judith S},
  journal={Human--computer interaction},
  volume={15},
  number={2-3},
  pages={139--178},
  year={2000},
  publisher={Taylor \& Francis}
}

@article{meyer2010rise,
  title={The rise of the knowledge broker},
  author={Meyer, Morgan},
  journal={Science communication},
  volume={32},
  number={1},
  pages={118--127},
  year={2010},
  publisher={Sage Publications Sage CA: Los Angeles, CA}
}

@article{linhares2022clinicalpath,
  title={Clinicalpath: a visualization tool to improve the evaluation of electronic health records in clinical decision-making},
  author={Linhares, Claudio DG and Lima, Daniel M and Ponciano, Jean R and Olivatto, Mauro M and Gutierrez, Marco A and Poco, Jorge and Traina, Caetano and Traina, Agma JM},
  journal={IEEE transactions on visualization and computer graphics},
  volume={29},
  number={10},
  pages={4031--4046},
  year={2022},
  publisher={IEEE}
}

@article{gui2020physician,
  title={Physician champions’ perspectives and practices on electronic health records implementation: challenges and strategies},
  author={Gui, Xinning and Chen, Yunan and Zhou, Xiaomu and Reynolds, Tera L and Zheng, Kai and Hanauer, David A},
  journal={JAMIA open},
  volume={3},
  number={1},
  pages={53--61},
  year={2020},
  publisher={Oxford University Press}
}

@inproceedings{cui2022automed,
  title={Automed: automated medical risk predictive modeling on electronic health records},
  author={Cui, Suhan and Wang, Jiaqi and Gui, Xinning and Wang, Ting and Ma, Fenglong},
  booktitle={2022 IEEE International Conference on Bioinformatics and Biomedicine (BIBM)},
  pages={948--953},
  year={2022},
  organization={IEEE}
}

@article{pine2022addressing,
  title={Addressing Fragmentation of Health Services through Data-Driven Knowledge Co-Production within a Boundary Organization},
  author={Pine, Kathleen H and Hinrichs, Margaret and Love, Kailey and Shafer, Michael and Runger, George and Riley, William},
  journal={The Journal of Community Informatics},
  volume={18},
  number={2},
  pages={3--26},
  year={2022}
}

@article{pine2023innovations,
  title={Innovations in clinical documentation integrity practice: Continual adaptation in a data-intensive healthcare organisation},
  author={Pine, Kathleen H and Landon, Lee Anne and Bossen, Claus and VanGelder, ME},
  journal={Health Information Management Journal},
  volume={52},
  number={2},
  pages={119--124},
  year={2023},
  publisher={SAGE Publications Sage UK: London, England}
}

@article{helou2019understanding,
  title={Understanding the situated roles of electronic medical record systems to enable redesign: Mixed methods study},
  author={Helou, Samar and Abou-Khalil, Victoria and Yamamoto, Goshiro and Kondoh, Eiji and Tamura, Hiroshi and Hiragi, Shusuke and Sugiyama, Osamu and Okamoto, Kazuya and Nambu, Masayuki and Kuroda, Tomohiro and others},
  journal={JMIR human factors},
  volume={6},
  number={3},
  pages={e13812},
  year={2019},
  publisher={JMIR Publications Inc., Toronto, Canada}
}

@inproceedings{pine2022investigating,
  title={Investigating Data Work Across Domains: New Perspectives on the Work of Creating Data},
  author={Pine, Kathleen and Bossen, Claus and Holten M{\o}ller, Naja and Miceli, Milagros and Lu, Alex Jiahong and Chen, Yunan and Horgan, Leah and Su, Zhaoyuan and Neff, Gina and Mazmanian, Melissa},
  booktitle={CHI Conference on Human Factors in Computing Systems Extended Abstracts},
  pages={1--6},
  year={2022}
}

@misc{bossen2019data,
  title={Data work in healthcare: An Introduction},
  author={Bossen, Claus and Pine, Kathleen H and Cabitza, Federico and Ellingsen, Gunnar and Piras, Enrico Maria},
  journal={Health Informatics Journal},
  volume={25},
  number={3},
  pages={465--474},
  year={2019},
  publisher={SAGE Publications Sage UK: London, England}
}

@article{rundo2020recent,
  title={Recent advances of HCI in decision-making tasks for optimized clinical workflows and precision medicine},
  author={Rundo, Leonardo and Pirrone, Roberto and Vitabile, Salvatore and Sala, Evis and Gambino, Orazio},
  journal={Journal of biomedical informatics},
  volume={108},
  pages={103479},
  year={2020},
  publisher={Elsevier}
}

@article{nagendran2020artificial,
  title={Artificial intelligence versus clinicians: systematic review of design, reporting standards, and claims of deep learning studies},
  author={Nagendran, Myura and Chen, Yang and Lovejoy, Christopher A and Gordon, Anthony C and Komorowski, Matthieu and Harvey, Hugh and Topol, Eric J and Ioannidis, John PA and Collins, Gary S and Maruthappu, Mahiben},
  journal={bmj},
  volume={368},
  year={2020},
  publisher={British Medical Journal Publishing Group}
}

@article{hernandez2020minimar,
  title={MINIMAR (MINimum Information for Medical AI Reporting): Developing reporting standards for artificial intelligence in health care},
  author={Hernandez-Boussard, Tina and Bozkurt, Selen and Ioannidis, John PA and Shah, Nigam H},
  journal={Journal of the American Medical Informatics Association},
  volume={27},
  number={12},
  pages={2011--2015},
  year={2020},
  publisher={Oxford University Press}
}

@inproceedings{singh2017hci,
  title={HCI and health: Learning from interdisciplinary interactions},
  author={Singh, Aneesha and Newhouse, Nikki and Gibbs, Jo and Blandford, Ann E and Chen, Yunan and Briggs, Pam and Mentis, Helena and Sellen, Kate M and Bardram, Jakob E},
  booktitle={Proceedings of the 2017 CHI Conference Extended Abstracts on Human Factors in Computing Systems},
  pages={1322--1325},
  year={2017}
}

@article{fossouo2023linking,
  title={Linking activity theory within user-centered design: novel framework to inform design and evaluation of adverse drug reaction reporting systems in pharmacy},
  author={Fossouo Tagne, Joel and Yakob, Reginald Amin and Mcdonald, Rachael and Wickramasinghe, Nilmini},
  journal={JMIR Human Factors},
  volume={10},
  pages={e43529},
  year={2023},
  publisher={JMIR Publications Toronto, Canada}
}

@inproceedings{park2012adaptation,
  title={Adaptation as design: learning from an EMR deployment study},
  author={Park, Sun Young and Chen, Yunan},
  booktitle={Proceedings of the SIGCHI Conference on Human Factors in Computing Systems},
  pages={2097--2106},
  year={2012}
}

@article{demir2012decision,
  title={A decision support tool for health service re-design},
  author={Demir, Eren and Chahed, Salma and Chaussalet, Thierry and Toffa, Sam and Fouladinajed, Farid},
  journal={Journal of medical systems},
  volume={36},
  pages={621--630},
  year={2012},
  publisher={Springer}
}

@book{engestrom2018expertise,
  title={Expertise in transition: Expansive learning in medical work},
  author={Engestr{\"o}m, Yrj{\"o}},
  year={2018},
  publisher={Cambridge University Press}
}

@book{nardi1996context,
  title={Context and consciousness: Activity theory and human--computer interaction},
  author={Nardi, BA},
  year={1996},
  publisher={The MIT Press}
}

@article{star1989institutional,
  title={Institutional ecology,translations' and boundary objects: Amateurs and professionals in Berkeley's Museum of Vertebrate Zoology, 1907-39},
  author={Star, Susan Leigh and Griesemer, James R},
  journal={Social studies of science},
  volume={19},
  number={3},
  pages={387--420},
  year={1989},
  publisher={Sage Publications London}
}

@article{saldana2021coding,
  title={Coding techniques for quantitative and mixed data},
  author={Salda{\~n}a, Johnny},
  journal={The Routledge reviewer’s guide to mixed methods analysis},
  pages={151--160},
  year={2021},
  publisher={Routledge}
}

@article{braun2019reflecting,
  title={Reflecting on reflexive thematic analysis},
  author={Braun, Virginia and Clarke, Victoria},
  journal={Qualitative research in sport, exercise and health},
  volume={11},
  number={4},
  pages={589--597},
  year={2019},
  publisher={Taylor \& Francis}
}

@article{braun2006using,
  title={Using thematic analysis in psychology},
  author={Braun, Virginia and Clarke, Victoria},
  journal={Qualitative research in psychology},
  volume={3},
  number={2},
  pages={77--101},
  year={2006},
  publisher={Taylor \& Francis}
}

@article{grundgeiger2024motives,
  title={Motives and fluent interaction in clinical documentation: An experience-based design for a documentation tool in anesthesiology},
  author={Grundgeiger, Tobias and Juranz, Christian and Hurtienne, J{\"o}rn and Happel, Oliver},
  journal={Human Factors in Healthcare},
  volume={5},
  pages={100071},
  year={2024},
  publisher={Elsevier}
}

@article{clemmensen2016making,
  title={Making HCI theory work: an analysis of the use of activity theory in HCI research},
  author={Clemmensen, Torkil and Kaptelinin, Victor and Nardi, Bonnie},
  journal={Behaviour \& Information Technology},
  volume={35},
  number={8},
  pages={608--627},
  year={2016},
  publisher={Taylor \& Francis}
}

@book{kaptelinin2012activity,
  title={Activity theory in HCI: Fundamentals and reflections},
  author={Kaptelinin, Victor and Nardi, Bonnie A},
  volume={13},
  year={2012},
  publisher={Morgan \& Claypool Publishers}
}

@incollection{durst2017guideline,
  title={A Guideline to Use Activity Theory for Collaborative Healthcare Information Systems Design},
  author={Durst, Carolin and Wickramasinghe, Nilmini and Riechert, Jana},
  booktitle={Handbook of Research on Healthcare Administration and Management},
  pages={616--626},
  year={2017},
  publisher={IGI Global}
}

@article{valecha2021activity,
  title={An activity theory approach to leak detection and mitigation in patient health information (PHI)},
  author={Valecha, Rohit and Upadhyaya, Shambhu and Rao, H Raghav},
  journal={Journal of the Association for Information Systems},
  volume={22},
  number={4},
  pages={6},
  year={2021}
}

@article{bjorn2009boundary,
  title={Boundary factors and contextual contingencies: configuring electronic templates for healthcare professionals},
  author={Bj{\o}rn, Pernille and Burgoyne, Sue and Crompton, Vicky and MacDonald, Teri and Pickering, Barbe and Munro, Sue},
  journal={European Journal of Information Systems},
  volume={18},
  number={5},
  pages={428--441},
  year={2009},
  publisher={Taylor \& Francis}
}

@article{jiang2023health,
  title={Health system-scale language models are all-purpose prediction engines},
  author={Jiang, Lavender Yao and Liu, Xujin Chris and Nejatian, Nima Pour and Nasir-Moin, Mustafa and Wang, Duo and Abidin, Anas and Eaton, Kevin and Riina, Howard Antony and Laufer, Ilya and Punjabi, Paawan and others},
  journal={Nature},
  volume={619},
  number={7969},
  pages={357--362},
  year={2023},
  publisher={Nature Publishing Group UK London}
}

@article{chen2020trends,
  title={Trends and features of the applications of natural language processing techniques for clinical trials text analysis},
  author={Chen, Xieling and Xie, Haoran and Cheng, Gary and Poon, Leonard KM and Leng, Mingming and Wang, Fu Lee},
  journal={Applied Sciences},
  volume={10},
  number={6},
  pages={2157},
  year={2020},
  publisher={MDPI}
}

@article{leontiev1978atividade,
  title={Atividade, consci{\^e}ncia e personalidade},
  author={Leontiev, Alexei N},
  journal={Buenos Aires: Ciencias del Hombre},
  year={1978}
}

@article{yamazaki2023exploration,
  title={Exploration of Interdisciplinary Fusion and Interorganizational Collaboration With the Advancement of AI Research: A Case Study on Natural Language Processing},
  author={Yamazaki, Tomomi and Sakata, Ichiro},
  journal={IEEE Transactions on Engineering Management},
  year={2023},
  publisher={IEEE}
}

@article{dourish2004we,
  title={What we talk about when we talk about context},
  author={Dourish, Paul},
  journal={Personal and ubiquitous computing},
  volume={8},
  pages={19--30},
  year={2004},
  publisher={Springer}
}

@article{janowski2023natural,
  title={Natural language processing techniques for clinical text analysis in healthcare},
  author={Janowski, Andrzej},
  journal={Journal of Advanced Analytics in Healthcare Management},
  volume={7},
  number={1},
  pages={51--76},
  year={2023}
}

@article{li2024academic,
  title={Academic collaboration on large language model studies increases overall but varies across disciplines},
  author={Li, Lingyao and Dinh, Ly and Hu, Songhua and Hemphill, Libby},
  journal={arXiv preprint arXiv:2408.04163},
  year={2024}
}

@article{iyengar2020big,
  title={Big data analytics in healthcare using spreadsheets},
  author={Iyengar, Samaya Pillai and Acharya, Haridas and Kadam, Manik},
  journal={Big Data Analytics in Healthcare},
  pages={155--187},
  year={2020},
  publisher={Springer}
}

@article{ponedal2002understanding,
  title={Understanding decision support systems},
  author={Ponedal, Steven},
  journal={Journal of Managed Care Pharmacy},
  volume={8},
  number={2},
  pages={96--101},
  year={2002},
  publisher={Academy of Managed Care Pharmacy}
}

@inproceedings{thorne2017role,
  title={The Role of Spreadsheets in Clinical Decision Support: A Survey of the Medical Algorithms Company User Community},
  author={Thorne, Simon},
  booktitle={18th EuSpRIG Annual Conference “Spreadsheet Risk Management”},
  volume={12},
  pages={137},
  year={2017}
}

@article{taylor2020research,
  title={Research skills and the data spreadsheet: A research primer for low-and middle-income countries},
  author={Taylor, David McD and Hodkinson, Peter W and Khan, Abdus Salam and Simon, Erin L},
  journal={African Journal of Emergency Medicine},
  volume={10},
  pages={S140--S144},
  year={2020},
  publisher={Elsevier}
}

@article{juluru2015use,
  title={Use of Spreadsheets for Research Data Collection and Preparation:: A Primer},
  author={Juluru, Krishna and Eng, John},
  journal={Academic radiology},
  volume={22},
  number={12},
  pages={1592--1599},
  year={2015},
  publisher={Elsevier}
}

@inproceedings{du2020try,
  title={" Try your best" parent behaviors during administration of an online language assessment tool for bilingual Mandarin-English children},
  author={Du, Yao and Sheng, Li and Tekinbas, Katie Salen},
  booktitle={proceedings of the interaction design and children conference},
  pages={409--420},
  year={2020}
}

@inproceedings{du2024voice,
  title={Voice Assistive Technology for Activities of Daily Living: Developing an Alexa Telehealth Training for Adults with Cognitive-Communication Disorders},
  author={Du, Yao and O'Connor, Claire and Byun, Ginna and Kim, Lauren H and Amrgousian, Siona and Vora, Priyal},
  booktitle={Proceedings of the CHI Conference on Human Factors in Computing Systems},
  pages={1--15},
  year={2024}
}

@article{leigh2010not,
  title={This is not a boundary object: Reflections on the origin of a concept},
  author={Leigh Star, Susan},
  journal={Science, technology, \& human values},
  volume={35},
  number={5},
  pages={601--617},
  year={2010},
  publisher={Sage Publications Sage CA: Los Angeles, CA}
}

@article{akkerman2011boundary,
  title={Boundary crossing and boundary objects},
  author={Akkerman, Sanne F and Bakker, Arthur},
  journal={Review of educational research},
  volume={81},
  number={2},
  pages={132--169},
  year={2011},
  publisher={Sage Publications Sage CA: Los Angeles, CA}
}

@article{caccamo2023boundary,
  title={Boundary objects, knowledge integration, and innovation management: A systematic review of the literature},
  author={Caccamo, Marta and Pittino, Daniel and Tell, Fredrik},
  journal={Technovation},
  volume={122},
  pages={102645},
  year={2023},
  publisher={Elsevier}
}

@article{carlile2002pragmatic,
  title={A pragmatic view of knowledge and boundaries: Boundary objects in new product development},
  author={Carlile, Paul R},
  journal={Organization science},
  volume={13},
  number={4},
  pages={442--455},
  year={2002},
  publisher={INFORMS}
}

@article{carlile2004transferring,
  title={Transferring, translating, and transforming: An integrative framework for managing knowledge across boundaries},
  author={Carlile, Paul R},
  journal={Organization science},
  volume={15},
  number={5},
  pages={555--568},
  year={2004},
  publisher={INFORMS}
}

@article{pawlowski2004bridging,
  title={Bridging user organizations: Knowledge brokering and the work of information technology professionals},
  author={Pawlowski, Suzanne D and Robey, Daniel},
  journal={MIS quarterly},
  pages={645--672},
  year={2004},
  publisher={JSTOR}
}

@book{tell2016managingknowledge,
    author = {Tell, Frederik and Berggren, Christian and Brusoni, Stefano and Van de Ven, Andrew},
    title = "{Managing Knowledge Integration Across Boundaries}",
    publisher = {Oxford University Press},
    year = {2016},
    month = {12},
    isbn = {9780198785972},
    doi = {10.1093/acprof:oso/9780198785972.001.0001},
    url = {https://doi.org/10.1093/acprof:oso/9780198785972.001.0001},
}

@article{van2017boundary,
  title={Boundary spanning, boundary objects, and innovation},
  author={Van de Ven, Andrew and Zahra, Shaker A},
  journal={Managing knowledge integration across boundaries},
  pages={241--254},
  year={2017},
  publisher={Oxford, UK: Oxford University Press}
}

@article{bowker1999sorting,
  title={Sorting things out},
  author={Bowker, Geoffrey and Star, Susan Leigh},
  journal={Classification and its consequences},
  volume={4},
  year={1999},
  publisher={Citeseer}
}

@inproceedings{reddy2001coordinating,
  title={Coordinating heterogeneous work: Information and representation in medical care},
  author={Reddy, Madhu C and Dourish, Paul and Pratt, Wanda},
  booktitle={ECSCW 2001: Proceedings of the Seventh European Conference on Computer Supported Cooperative Work 16--20 September 2001, Bonn, Germany},
  pages={239--258},
  year={2001},
  organization={Springer}
}

@article{wilson2007boundary,
  title={Boundary objects as rhetorical exigence: Knowledge mapping and interdisciplinary cooperation at the Los Alamos National Laboratory},
  author={Wilson, Greg and Herndl, Carl G},
  journal={Journal of Business and Technical Communication},
  volume={21},
  number={2},
  pages={129--154},
  year={2007},
  publisher={Sage Publications Sage CA: Thousand Oaks, CA}
}

@article{lomas2007between,
  title={The in-between world of knowledge brokering},
  author={Lomas, Jonathan},
  journal={Bmj},
  volume={334},
  number={7585},
  pages={129--132},
  year={2007},
  publisher={British Medical Journal Publishing Group}
}

@article{friman2010understanding,
  title={Understanding boundary work through discourse theory: Inter/disciplines and interdisciplinarity},
  author={Friman, Mathias},
  journal={Science \& Technology Studies},
  volume={23},
  number={2},
  pages={5--19},
  year={2010}
}

@techreport{blackwell2009radical,
  title={Radical innovation: crossing knowledge boundaries with interdisciplinary teams},
  author={Blackwell, Alan F and Wilson, Lee and Boulton, Charles and Knell, John},
  year={2009},
  institution={University of Cambridge, Computer Laboratory}
}

@article{barry2008logics,
  title={Logics of interdisciplinarity},
  author={Barry, Andrew and Born, Georgina and Weszkalnys, Gisa},
  journal={Economy and society},
  volume={37},
  number={1},
  pages={20--49},
  year={2008},
  publisher={Taylor \& Francis}
}

@article{schmidt2008towards,
  title={Towards a philosophy of interdisciplinarity: An attempt to provide a classification and clarification},
  author={Schmidt, Jan C},
  journal={Poiesis \& Praxis},
  volume={5},
  pages={53--69},
  year={2008},
  publisher={Springer}
}

@article{klein2010taxonomy,
  title={A taxonomy of interdisciplinarity},
  author={Klein, Julie Thompson},
  journal={The Oxford handbook of interdisciplinarity},
  volume={15},
  number={6},
  pages={15},
  year={2010}
}

@inproceedings{zhou2011cpoe,
  title={CPOE workarounds, boundary objects, and assemblages},
  author={Zhou, Xiaomu and Ackerman, Mark and Zheng, Kai},
  booktitle={Proceedings of the SIGCHI Conference on Human Factors in Computing Systems},
  pages={3353--3362},
  year={2011}
}

@inproceedings{agapie2024conducting,
  title={Conducting Research at the Intersection of HCI and Health: Building and Supporting Teams with Diverse Expertise to Increase Public Health Impact},
  author={Agapie, Elena and Karkar, Ravi and Aung, Tricia and Burgess, Eleanor R and Chinguwa, Munyaradzi Joel and Graham, Andrea K and Klasnja, Predrag and Lyon, Aaron and McCall, Terika and Munson, Sean A and others},
  booktitle={Extended Abstracts of the CHI Conference on Human Factors in Computing Systems},
  pages={1--6},
  year={2024}
}

@inproceedings{lyon2023bridging,
  title={Bridging HCI and Implementation Science for Innovation Adoption and Public Health Impact},
  author={Lyon, Aaron and Munson, Sean A and Reddy, Madhu and Schueller, Stephen M and Agapie, Elena and Yarosh, Svetlana and Dopp, Alex and von Thiele Schwarz, Ulrica and Doherty, Gavin and Graham, Andrea K and others},
  booktitle={Extended Abstracts of the 2023 CHI Conference on Human Factors in Computing Systems},
  pages={1--7},
  year={2023}
}

@inproceedings{bednarik2022integration,
  title={Integration of human factors in surgery: Interdisciplinary collaboration in design, development, and evaluation of surgical technologies},
  author={Bednarik, Roman and Blandford, Ann and Feng, Feng and Huotarinen, Antti and Iso-Mustaj{\"a}rvi, Matti and Lee, Ahreum and Nicolosi, Federico and Opie, Jeremy and Yoo, Soojeong and Zheng, Bin},
  booktitle={CHI conference on human factors in computing systems extended abstracts},
  pages={1--7},
  year={2022}
}

@inproceedings{mozgai2024accelerating,
  title={Accelerating Scoping Reviews: A Case Study in the User-Centered Design of an AI-Enabled Interdisciplinary Research Tool},
  author={Mozgai, Sharon A and Kaurloto, Cari and Winn, Jade G and Leeds, Andrew and Beland, Sarah and Sookiassian, Arman and Hartholt, Arno},
  booktitle={Extended Abstracts of the CHI Conference on Human Factors in Computing Systems},
  pages={1--8},
  year={2024}
}

@article{engestrom2000activity,
  title={Activity theory as a framework for analyzing and redesigning work},
  author={Engestrom, Yrjo},
  journal={Ergonomics},
  volume={43},
  number={7},
  pages={960--974},
  year={2000},
  publisher={Taylor \& Francis}
}

@article{kimble2010innovation,
  title={Innovation and knowledge sharing across professional boundaries: Political interplay between boundary objects and brokers},
  author={Kimble, Chris and Grenier, Corinne and Goglio-Primard, Karine},
  journal={International journal of information management},
  volume={30},
  number={5},
  pages={437--444},
  year={2010},
  publisher={Elsevier}
}

@article{bharosa2012activity,
  title={An activity theory analysis of boundary objects in cross-border information systems development for disaster management},
  author={Bharosa, Nitesh and Lee, JinKyu and Janssen, Marijn and Rao, H Raghav},
  journal={Security Informatics},
  volume={1},
  pages={1--17},
  year={2012},
  publisher={Springer}
}

@article{mao2019data,
  title={How data scientistswork together with domain experts in scientific collaborations: To find the right answer or to ask the right question?},
  author={Mao, Yaoli and Wang, Dakuo and Muller, Michael and Varshney, Kush R and Baldini, Ioana and Dugan, Casey and Mojsilovi{\'c}, Aleksandra},
  journal={Proceedings of the ACM on Human-Computer Interaction},
  volume={3},
  number={GROUP},
  pages={1--23},
  year={2019},
  publisher={ACM New York, NY, USA}
}

@article{hou2017hacking,
  title={Hacking with NPOs: collaborative analytics and broker roles in civic data hackathons},
  author={Hou, Youyang and Wang, Dakuo},
  journal={Proceedings of the ACM on Human-Computer Interaction},
  volume={1},
  number={CSCW},
  pages={1--16},
  year={2017},
  publisher={ACM New York, NY, USA}
}

@book{spinuzzi2008network,
  title={Network: Theorizing knowledge work in telecommunications},
  author={Spinuzzi, Clay},
  year={2008},
  publisher={Cambridge University Press}
}

@book{kaptelinin2009acting,
  title={Acting with technology: Activity theory and interaction design},
  author={Kaptelinin, Victor and Nardi, Bonnie A},
  year={2009},
  publisher={MIT press}
}

@book{engestrom2015learning,
  title={Learning by expanding},
  author={Engestr{\"o}m, Yrj{\"o}},
  year={2015},
  publisher={Cambridge University Press}
}

@article{yang2025recover,
  title={RECOVER: Designing a Large Language Model-based Remote Patient Monitoring System for Postoperative Gastrointestinal Cancer Care},
  author={Yang, Ziqi and Lu, Yuxuan and Bagdasarian, Jennifer and Swain, Vedant Das and Agarwal, Ritu and Campbell, Collin and Al-Refaire, Waddah and El-Bayoumi, Jehan and Gao, Guodong and Wang, Dakuo and others},
  journal={arXiv preprint arXiv:2502.05740},
  year={2025}
}

@article{yao2025more,
  title={More Modality, More AI: Exploring Design Opportunities of AI-Based Multi-modal Remote Monitoring Technologies for Early Detection of Mental Health Sequelae in Youth Concussion Patients},
  author={Yao, Bingsheng and Zhao, Menglin and Sun, Yuling and Cao, Weidan and Yin, Changchang and Intille, Stephen and Xu, Xuhai and Zhang, Ping and Yang, Jingzhen and Wang, Dakuo},
  journal={arXiv preprint arXiv:2502.03732},
  year={2025}
}

@article{wu2024clinical,
  title={Clinical Challenges and AI Opportunities in Decision-Making for Cancer Treatment-Induced Cardiotoxicity},
  author={Wu, Siyi and Cao, Weidan and Fu, Shihan and Yao, Bingsheng and Yang, Ziqi and Yin, Changchang and Mishra, Varun and Addison, Daniel and Zhang, Ping and Wang, Dakuo},
  journal={arXiv preprint arXiv:2408.03586},
  year={2024}
}

@article{wu2024cardioai,
  title={CardioAI: A Multimodal AI-based System to Support Symptom Monitoring and Risk Detection of Cancer Treatment-Induced Cardiotoxicity},
  author={Wu, Siyi and Cao, Weidan and Fu, Shihan and Yao, Bingsheng and Yang, Ziqi and Yin, Changchang and Mishra, Varun and Addison, Daniel and Zhang, Ping and Wang, Dakuo},
  journal={arXiv preprint arXiv:2410.04592},
  year={2024}
}

@article{yang2024talk2care,
  title={Talk2Care: An LLM-based Voice Assistant for Communication between Healthcare Providers and Older Adults},
  author={Yang, Ziqi and Xu, Xuhai and Yao, Bingsheng and Rogers, Ethan and Zhang, Shao and Intille, Stephen and Shara, Nawar and Gao, Guodong Gordon and Wang, Dakuo},
  journal={Proceedings of the ACM on Interactive, Mobile, Wearable and Ubiquitous Technologies},
  volume={8},
  number={2},
  pages={1--35},
  year={2024},
  publisher={ACM New York, NY, USA}
}

@article{xu2024mental,
  title={Mental-llm: Leveraging large language models for mental health prediction via online text data},
  author={Xu, Xuhai and Yao, Bingsheng and Dong, Yuanzhe and Gabriel, Saadia and Yu, Hong and Hendler, James and Ghassemi, Marzyeh and Dey, Anind K and Wang, Dakuo},
  journal={Proceedings of the ACM on Interactive, Mobile, Wearable and Ubiquitous Technologies},
  volume={8},
  number={1},
  pages={1--32},
  year={2024},
  publisher={ACM New York, NY, USA}
}

@inproceedings{bardram2011activity,
  title={Activity analysis: applying activity theory to analyze complex work in hospitals},
  author={Bardram, Jakob and Doryab, Afsaneh},
  booktitle={Proceedings of the ACM 2011 conference on Computer supported cooperative work},
  pages={455--464},
  year={2011}
}

@article{bardram2009activity,
  title={Activity-based computing for medical work in hospitals},
  author={Bardram, Jakob E},
  journal={ACM Transactions on Computer-Human Interaction (TOCHI)},
  volume={16},
  number={2},
  pages={1--36},
  year={2009},
  publisher={ACM New York, NY, USA}
}

@article{kerosuo2010promoting,
  title={Promoting innovation and learning through Change Laboratory: An example from Finnish Health care},
  author={Kerosuo, Hannele and Kajamaa, Anu and Engestr{\"o}m, Yrj{\"o}},
  journal={Central European Journal of Public Policy},
  volume={4},
  number={1},
  pages={110--131},
  year={2010},
  publisher={Center for Social and Economic Strategies, Faculty of Social Sciences~…}
}

@article{almalki2016activity,
  title={Activity theory as a theoretical framework for health self-quantification: a systematic review of empirical studies},
  author={Almalki, Manal and Gray, Kathleen and Martin-Sanchez, Fernando},
  journal={Journal of medical Internet research},
  volume={18},
  number={5},
  pages={e131},
  year={2016},
  publisher={JMIR Publications Toronto, Canada}
}

@article{schmidt1992taking,
  title={Taking CSCW seriously: Supporting articulation work},
  author={Schmidt, Kjeld and Bannon, Liam},
  journal={Computer supported cooperative work (CSCW)},
  volume={1},
  number={1},
  pages={7--40},
  year={1992},
  publisher={Springer}
}

@article{assistance2003summary,
  title={Summary of the hipaa privacy rule},
  author={Assistance, HIPAA Compliance},
  journal={Office for Civil Rights},
  year={2003}
}

@inproceedings{bannon1997constructing,
  title={Constructing common information spaces},
  author={Bannon, Liam and B{\o}dker, Susanne},
  booktitle={Proceedings of the Fifth European Conference on Computer Supported Cooperative Work},
  pages={81--96},
  year={1997},
  organization={Springer}
}

@inproceedings{amershi2019guidelines,
  title={Guidelines for human-AI interaction},
  author={Amershi, Saleema and Weld, Dan and Vorvoreanu, Mihaela and Fourney, Adam and Nushi, Besmira and Collisson, Penny and Suh, Jina and Iqbal, Shamsi and Bennett, Paul N and Inkpen, Kori and others},
  booktitle={Proceedings of the 2019 chi conference on human factors in computing systems},
  pages={1--13},
  year={2019}
}

@inproceedings{bossen2002parameters,
  title={The parameters of common information spaces: The heterogeneity of cooperative work at a hospital ward},
  author={Bossen, Claus},
  booktitle={Proceedings of the 2002 ACM conference on Computer supported cooperative work},
  pages={176--185},
  year={2002}
}

@article{furnas1987vocabulary,
  title={The vocabulary problem in human-system communication},
  author={Furnas, George W. and Landauer, Thomas K. and Gomez, Louis M. and Dumais, Susan T.},
  journal={Communications of the ACM},
  volume={30},
  number={11},
  pages={964--971},
  year={1987},
  publisher={ACM New York, NY, USA}
}

@article{collins2007trading,
  title={Trading zones and interactional expertise},
  author={Collins, Harry and Evans, Robert and Gorman, Mike},
  journal={Studies in History and Philosophy of Science Part A},
  volume={38},
  number={4},
  pages={657--666},
  year={2007},
  publisher={Elsevier}
}

@article{bechky2003sharing,
  title={Sharing meaning across occupational communities: The transformation of understanding on a production floor},
  author={Bechky, Beth A},
  journal={Organization science},
  volume={14},
  number={3},
  pages={312--330},
  year={2003},
  publisher={INFORMS}
}

@article{orlikowski1994technological,
  title={Technological frames: making sense of information technology in organizations},
  author={Orlikowski, Wanda J and Gash, Debra C},
  journal={ACM Transactions on Information Systems (TOIS)},
  volume={12},
  number={2},
  pages={174--207},
  year={1994},
  publisher={ACM New York, NY, USA}
}

@inproceedings{star1994steps,
  title={Steps towards an ecology of infrastructure: complex problems in design and access for large-scale collaborative systems},
  author={Star, Susan Leigh and Ruhleder, Karen},
  booktitle={Proceedings of the 1994 ACM conference on Computer supported cooperative work},
  pages={253--264},
  year={1994}
}

@article{schmidt1994cooperative,
  title={Cooperative work and its articulation: requirements for computer support},
  author={Schmidt, Kjeld},
  journal={Le travail humain},
  pages={345--366},
  year={1994},
  publisher={JSTOR}
}

@book{galison1997image,
  title={Image and logic: A material culture of microphysics},
  author={Galison, Peter},
  year={1997},
  publisher={University of Chicago Press}
}

@book{suchman1987plans,
  title={Plans and situated actions: The problem of human-machine communication},
  author={Suchman, Lucille Alice},
  year={1987},
  publisher={Cambridge university press}
}

\newpage
\appendix

\section{Interview Protocol}
\label{app:protocol}

We tailor the interview protocol to the unique roles of clinical experts and technical experts.

\textbf{Interview Protocol For Clinical Experts}

\textit{Part I. Clinical Experts' Use of Non-EMR Tools in Clinical Practice}

\begin{enumerate}
    \item Can you briefly describe the clinical setting of the interdisciplinary project [VAT Alexa project (tele-rehabilitation for cognitive-communication disorders) OR Bilingual MERLS project (bilingual telehealth language assessment)] you are currently involved in? 
    \item When you're documenting OR analyzing clinical information, what specific purposes or goals are you aiming to achieve? (Object) 
    \item What types of tools, other than EMRs, do you use to document clinical information in your practice? (Tools)
    \item Is there any guidance, whether formal or informal, like policies or protocols, that affects how you use such tools? (Rules) 
    \begin{enumerate}
        \item Who sets these guidelines, and how do you go about following them?
    \end{enumerate}
    \item How do you use these tools to meet your clinical goals and documentation needs? (Tools)
    \begin{enumerate}
        \item What kind of information do you document in these tools, and what are the reasons behind this?
        \item Which features of these tools do you find most helpful in your clinical work? (Outcome)
        \begin{enumerate}
            \item Could you share why these features are particularly useful to you?
        \end{enumerate}
        \item Are there any challenges you encounter when using such tools? (Outcome)
        \begin{enumerate}
            \item How do you typically deal with these challenges?
        \end{enumerate}
    \end{enumerate}
    \item Do you collaborate with others when using such tools? (Community) 
    \begin{enumerate}
        \item If so, how does that collaboration usually work? 
    \end{enumerate}
    \item How are tasks and responsibilities shared among you and your colleagues when it comes to using these tools for documentation? (Division of Labor) 

\end{enumerate}

\textit{Part II. Integration and Adaptation of non-EMR tools in Clinical NLP Research Collaboration}

Imagine you are explaining how you use spreadsheets for project MERLS or VAT to an NLP researcher. The researcher might not have clinical knowledge, but they understand the basics of the clinical setting for this project.

\begin{enumerate}
    \item How do you see your role in relation to the NLP researchers? Do you consider yourself a guide, a collaborator, a learner, or something else? (Subject)
    \begin{enumerate}
        \item What's your understanding of NLP technology as it applies to your work?
    \end{enumerate}
    
    \item How do you use tools, such as spreadsheets, in your collaboration with NLP researchers? (Tools)
    \begin{enumerate}
        \item Could you give some examples of how these tools help you with your tasks? [feel free to open your spreadsheet or documents and share your screen here to help you explain details as needed]
        \begin{enumerate}
            \item How would you describe your clinical setting (including the use of non-EMR tools) to an NLP researcher?
            \item How would you describe the types of information you work with, including any specific terminology, to an NLP researcher?
        \end{enumerate}
    \end{enumerate}

    \item Are there any limitations or challenges you face with these tools when working towards your collaborative goals? (Contradictions and Challenges)
    \item \textbf{Alternative Question} - Have you noticed any contradictions or tensions in your collaboration with NLP researchers (e.g., different goals, communication barriers)? (Contradictions and Challenges)
    \begin{enumerate}
        \item How have these contradictions been managed or resolved?
        \item Which part(s) of the clinical process do you think would be most difficult for NLP researchers to understand?
    \end{enumerate}
    \item How do you and the NLP researchers divide tasks and responsibilities in your project? (Division of Labor)
    \item Are there any changes or improvements you'd like to see, with or without AI, to better support future collaborations with NLP researchers?
    \item What additional features or functionalities would you like to have in non-EMR tools to make your collaboration more effective? (Tools/Artifacts Improvement)

\end{enumerate}

\textbf{Interview Protocol For Technical Experts}

\textit{Part I. Technical Experts' Understanding of Clinical Experts' Use of Non-EMR Data/Tools in Clinical Practice}

\begin{enumerate}
    \item Can you briefly describe the clinical aims of the interdisciplinary project Bilingual MERLS project (bilingual telehealth language assessment)] you are currently involved in? 
    \item As an NLP researcher, when you're analyzing/using clinical information, what specific purposes or goals are you aiming to achieve? (Object) 
    \item What types of tools did clinicians use to document clinical information in this project? (Tools)
    \item Were you given any clinical background knowledge of how clinicians organize and document the information using such tools (e.g., spreadsheet)  (Rules) 
    \begin{enumerate}
        \item Who (may be able to) provide this clinical background knowledge? 
    \end{enumerate}
    \item As an NLP researcher, do you think you understand the information given to you via such tools (e.g., spreadsheets)? 
    \begin{enumerate}
        \item Which features of these tools do you find most helpful in understanding clinical information? (Outcome)
        \begin{enumerate}
            \item Could you share why these features are particularly useful to you?
        \end{enumerate}
        \item Are there any challenges in understanding the information given to you using such tools? (Outcome)
        \begin{enumerate}
            \item How do you typically deal with these challenges?
        \end{enumerate}
    \end{enumerate}
    \item As an NLP researcher, how do you use this tool to conduct your NLP research? (Tools)
    \begin{enumerate}
        \item Which features of these tools do you find most helpful in conducting your NLP research? (Outcome)
        \begin{enumerate}
            \item Could you share why these features are particularly useful to you?
        \end{enumerate}
        \item Are there any challenges in conducting your NLP research? (Outcome)
        \begin{enumerate}
            \item How do you typically deal with these challenges?
        \end{enumerate}
    \end{enumerate}

    \item Do you collaborate with others when using such tools? (Community) 
    \begin{enumerate}
        \item If so, how does that collaboration usually work? 
    \end{enumerate}
    \item How are tasks and responsibilities shared among you and your colleagues when it comes to analyzing these clinical data for NLP research? (Division of Labor) 
    \begin{enumerate}
        \item How often have you met as a team? How many hours do you individual / collaborate?
    \end{enumerate}
    
\end{enumerate}

\textit{Part II. Integration and Adaptation of non-EMR tools in Clinical NLP Research Collaboration}

Imagine you are talking to clinicians who are using spreadsheets as source data for your NLP development in an interdisciplinary collaboration. The clinicians might not have NLP knowledge, but they understand the clinical setting for this project. 

\begin{enumerate}
    \item In your language, can you describe what the contribution of the NLP research is, and how it applies to clinician’s existing clinical work? 
    \item How do you see your role in relation to the clinicians? Do you consider yourself a guide, a collaborator, a learner, or teacher, something else when you are working together with clinicians? (Subject) 
    \item How would you describe your NLP research (including the use of non-EMR tools) to a clinician?
    \begin{enumerate}
        \item How would you describe the types of information you work with, including any specific terminology, to a clinician?
        \item What are the differences between different types of data (text vs. numerical data)? Can you share what type of information are being captured in text vs. numbers?
    \end{enumerate}
    \item Are there any limitations or challenges you face with these tools when working with clinicians? (Contradictions and Challenges)
    \item Alternative Question - Have you noticed any tensions in your collaboration with clinicians (e.g., different goals, communication barriers)? (Contradictions and Challenges)
    \item Which part(s) of the NLP process do you think would be most difficult for clinicians to understand?
    \item How do you address these difficulties during the collaboration with clinicians? 
    \item How do you and the clinicians collaboratively solve the difficulties in your research project? (Division of Labor)
    \item Are there any changes or improvements you'd like to see, with or without AI/NLP, to better support your future collaborations with clinicians?
    \item What additional features or functionalities would you like to have in these tools to make your collaboration more effective? (Tools/Artifacts Improvement)
    \item Is there anything you would like to learn from clinicians that can better facilitate collaboration and research development?

\end{enumerate}

    

\section{Codebook}
\label{app:codebook}
\begin{table*}[b]
\label{tab:codebook}
\begin{tabularx}{\textwidth}{@{}>{\raggedright\arraybackslash}p{2.8cm} >{\raggedright\arraybackslash}X >{\raggedright\arraybackslash}p{4.2cm}@{}}
\toprule
\textbf{Themes} & \textbf{Definitions} & \textbf{Sample Codes} \\ \midrule
Subject & Individual or subgroup chosen as the point of view in the analysis & Clinician (supervisor) \\
Community & Individuals or subgroups who share the same general object. & Graduate student clinicians / other clinicians \\
Tools (Instruments) & Physical or psychological artifacts (e.g., language, writing, software, etc.) & Physical; Digital; Psychological \\
Object & The “raw material” or “problem space” at which the activity is directed and which is molded or transformed into “outcomes” & Patient; Patient Data \\
Outcome & The goal/purpose for the object & Clinical outcomes \\
Rules & Explicit/implicit regulations, norms, conventions that constrain action/interaction & Explicit Rule; Implicit Rule \\
Activities & Clinical and technical activities within and across clinical vs.\ NLP communities & Clinical activities; Technical activities \\
\bottomrule
\end{tabularx}
\caption{Codebook for data analysis.}
\end{table*}

\end{document}